\pdfoutput=1

\documentclass[11pt,a4paper]{article}
\usepackage{jheppub}

\usepackage{amsmath}
\usepackage{amssymb}
\usepackage{graphicx,color,slashed,braket}
\usepackage{bbm}
\usepackage{ifpdf}
\usepackage{comment}

\usepackage{float}
\usepackage{hhline}
\usepackage{soul}

\usepackage{adjustbox}

\usepackage{empheq}

\newcommand{\be}{\begin{equation}}
\newcommand{\ee}{\end{equation}}
\newcommand{\ba}{\begin{eqnarray}}
\newcommand{\ea}{\end{eqnarray}}
\newcommand{\nn}{\nonumber}
\newcommand{\lp}{\left(}
\newcommand{\rp}{\right)}

\newcommand{\w}{\wedge}

\newcommand{\beq}{\begin{equation}}
\newcommand{\eeq}{\end{equation}}
\def\bea#1\eea{\begin{align}#1\end{align}}
\def\beal#1\eeal{\begin{subequations}\begin{align}#1\end{align}\end{subequations}}

\def\del {\partial}
\def\d {{\rm d}}

\title{Automated consistent truncations and stability \hspace{1in} of flux compactifications}

\author[a]{David Andriot,}
\author[b]{Paul Marconnet,}
\author[c]{Muthusamy Rajaguru,}
\author[c]{and Timm Wrase}

\affiliation[a]{Laboratoire d’Annecy-le-Vieux de Physique Th\'eorique (LAPTh),\\
CNRS, Universit\'e Savoie Mont Blanc (USMB), UMR 5108\\
9 Chemin de Bellevue, 74940 Annecy, France\\}
\affiliation[b]{Institut de Physique des 2 Infinis de Lyon,\\
Universit\'{e} de Lyon, UCBL, UMR 5822, CNRS/IN2P3,\\
4 rue Enrico Fermi, 69622 Villeurbanne Cedex, France\\}
\affiliation[c]{Department of Physics, Lehigh University, \\16 Memorial Drive East, Bethlehem, PA 18018, USA\\}

\emailAdd{andriot@lapth.cnrs.fr}
\emailAdd{marconnet@ipnl.in2p3.fr}
\emailAdd{muthusamy.rajaguru@lehigh.edu}
\emailAdd{timm.wrase@lehigh.edu}

\abstract{

\noindent
Classical flux compactifications contribute to a well-controlled corner of the string landscape, therefore providing an important testing ground for a variety of conjectures. We focus here on type II supergravity compactifications on 6d group manifolds towards 4d maximally symmetric spacetimes. We develop a code where the truncation to left-invariant scalars and the dimensional reduction to a 4d theory are automated, for any possible configuration of $O_p$-planes and $D_p$-branes. We then prove that any such truncation is consistent. We further compute the mass spectrum and analyse the stability of many de Sitter, Minkowski or anti-de Sitter solutions, as well as their consistency with swampland conjectures.
}

\begin{document}

\maketitle



\section{Introduction}\label{sec:introduction}

\indent

String theory backgrounds with maximally symmetric spacetimes, namely de Sitter, Minkowski or anti-de Sitter solutions, are ubiquitous in string phenomenology, holography and their ramifications. Classifying these solutions and understanding their properties therefore drives a lot of activity. Related conjectures \cite{Danielsson:2018ztv, Obied:2018sgi, Gautason:2018gln, Lust:2019zwm, Andriot:2022yyj} have appeared in the context of the swampland program \cite{Vafa:2005ui, Palti:2019pca}. Of particular interest is the question of stability of these solutions: non-supersymmetric solutions with maximally symmetric spacetimes have been conjectured to be unstable: see e.g.~\cite{Andriot:2018wzk, Garg:2018reu, Ooguri:2018wrx, Andriot:2018mav} for de Sitter, \cite{Acharya:2019mcu, Acharya:2020hsc} for Minkowski and \cite{Ooguri:2016pdq} for anti-de Sitter. Those conjectures are subject to many tests in the literature; some of them are also challenged by existing constructions in string theory settings. In this paper, we develop numerical tools to analyse the perturbative stability of classical flux compactifications, and apply them on a database of such solutions.\\

We focus in this work on solutions of 10-dimensional (10d) type IIA/B supergravities, with 4d maximally symmetric spacetimes. Those are candidates to be classical string backgrounds; whether or not they are in the classical string regime remains to be verified, but this does not affect the results of this paper. For these 10d solutions, we follow the ansatz and conventions of \cite{Andriot:2022way}: the 6d space is a group manifold (whose compactness is a priori not ensured, but analysed in detail in \cite{Andriot:2022yyj}), the fluxes have constant components in the left-invariant basis, and the only extended objects included are $D_p$-branes and orientifold $O_p$-planes. The latter are smeared, going together with a constant dilaton and no warp factor. This common ansatz is discussed in more detail in Section \ref{sec:review}. While this seems at first sight restricted, let us emphasize that this ansatz allows for a large variety of solutions: (non)-supersymmetric, (un)stable, (non)-scale separated, etc. These solutions may also be extended beyond this ansatz, in particular towards localized versions \cite{Andriot:2016xvq, Junghans:2020acz, Marchesano:2020qvg, Cribiori:2021djm, Marchesano:2022rpr}.

Such solutions have been classified in \cite{Andriot:2022way} according to their $O_p/D_p$ sources. In this classification, a first distinction to be made is whether sources have only one dimensionality $p$ (denoted by $s$ for single) or several ones (denoted by $m$ for multiple). A second crucial information is the number of different sets of directions wrapped by the $O_p$-planes. This has indeed important implications for the corresponding orientifold projections. For instance, a configuration with $O_5$ along internal (6d) directions 1 and 2, and further $O_5$ along directions 34, would be in the solution class $s_{55}$. If the source configuration contains additional $D_7$ but no further orientifold, then the class is $m_{55}$. We finally distinguish solutions with de Sitter, Minkowski or anti-de Sitter 4d spacetime by the sign of their cosmological constant, that we indicate respectively as $s_{55}^+$, $s_{55}^0$ or $s_{55}^-$. Overall, 21 solution classes have been identified in \cite{Andriot:2022way}, and filled with a list of known solutions. In particular, this was the case of de Sitter solutions $s_{55}^+ 1-17$ of \cite{Andriot:2020wpp} and $s_{55}^+ 18-27$ of \cite{Andriot:2021rdy}. New solutions were found in \cite{Andriot:2022way} thanks to the code {\tt MaxSymSolSearch} ({\tt MSSS}). These solutions completed the previous ones into a database made available with \cite{Andriot:2022way}, that we will use here.\\

Studying the stability of these 10d solutions is typically done by performing a dimensional reduction to a 4d theory of the form
\be\label{eq:4dactionIntro}
S_{4d} = \int d^4x \sqrt{-g}\lp \frac{M_p^2}{2} R_4 -\frac12 K_{ij}\partial_\mu \varphi^i \partial^\mu \varphi^j - V(\varphi^i)\rp\,.
\ee
It describes 4d scalar fields $\varphi^i$ minimally coupled to gravity and subject to a scalar potential $V$. These fields can be understood as fluctuations around the 10d solutions, and as such they capture some of their (in)stability. More precisely, the 10d solutions will turn out to correspond to critical points of the potential (loosely referred to as extrema in the following), i.e.~points in field space where $\del_{\varphi^i} V=0 \  \forall i$. The stability is then captured by the (sign of the) eigenvalues of the Hessian of the potential, related to those of the mass matrix, as we will explain. This led in \cite{Andriot:2022yyj} to analyse the stability of solutions by considering a restricted set of fields, $(\rho, \tau, \sigma_I)$, first introduced in \cite{Danielsson:2012et}, corresponding to some combinations of diagonal metric fluctuations and the dilaton. This analysis had been automated and performed with the code {\tt MaxSymSolSpec} ({\tt MSSSp}).

As in any dimensional reduction, a fundamental aspect is the truncation: the 10d fields, developed on an infinite basis of 6d modes, need to be truncated to a finite set, whose physics will be described by the 4d theory. There exist different choices of inequivalent truncations. Phenomenologically, the most relevant one is a low energy truncation: one truncates to the lightest modes. In practice, this is difficult to realise since it requires to first determine the complete mass spectrum, in order to identify the lightest modes. What is however often considered is the truncation to massless scalar fields, a.k.a. moduli, which are simpler to determine. Another common truncation is called a consistent truncation: the corresponding 4d theory describing the resulting finite set of modes is such that any of its 4d solutions also solves the 10d equations of motion. In practice, this corresponds to a finite set of modes which are, in some sense, independent or decoupled from those truncated. This set of modes may however contain both light and heavy modes, while other light modes may have been truncated. In this work, we will perform consistent truncations: as we will recall, finding an instability within this set of modes is sufficient to conclude on the instability of the 10d solution. Although phenomenologically debatable, this truncation will then be enough for our purposes. We implement it, as well as the corresponding dimensional reduction and resulting 4d theory, in an automated fashion in the code {\tt MSSV}.

Proving that a truncation is consistent is challenging. It first requires to find an appropriate truncation ansatz, and then verify that all 4d equations are captured by 10d equations. As we will review in Section \ref{sec:trunc}, it remains expected that the truncation of the 10d fields to left-invariant fluctuations on group manifolds is a consistent truncation, even in presence of (smeared) $O_p/D_p$ sources. The resulting theory is expected to be a 4d gauged supergravity. But this has typically been verified in a case by case analysis, for various compactifications and source configurations. In this work, thanks to a detailed comparison between the 10d equations given by the code {\tt MSSS} \cite{Andriot:2022way} and the 4d equations provided by the new code {\tt MSSV}, we prove for 4d maximally symmetric spacetimes that the truncation to left-invariant scalar fields on group manifolds is consistent for all 21 solution classes of \cite{Andriot:2022way} in type IIA/B, corresponding to various $O_p/D_p$ source configurations. We actually get more: we show a perfect matching between 10d and (combinations of) 4d equations of motion, with the same amount of equations on both sides. This means that there is actually no extra degree of freedom in the 10d ansatz (even though there are more constraints to satisfy in 10d). This matching ensures that the 10d solutions of \cite{Andriot:2020wpp, Andriot:2021rdy, Andriot:2022way} are critical points of our 4d scalar potential. This will allow us to study their stability using the 4d theory, and corresponding tools in {\tt MSSV}. Note that as usual, 10d Bianchi identities (in particular tadpole cancelation conditions) are not reproduced by the 4d theory and come as extra constraints when looking for solutions.\\

The paper is organised as follows. We first discuss in Section \ref{sec:trunc} consistent truncations and truncations to left-invariant modes on group manifolds, before specifying our truncation ansatz, and recalling the orientifold projections. We then detail in Section \ref{sec:dimred} the dimensional reduction, starting from 10d type II supergravities and compactifying towards a 4d maximally symmetric spacetime, ending up with a 4d theory of the form \eqref{eq:4dactionIntro}. We give in particular the scalar potential including the axions in equations \eqref{VIIA} and \eqref{VIIB}. We also discuss the computation of the scalar fields’ kinetic terms. We can then motivate and define the mass matrix and the $\eta_V$ parameter, to be used in stability studies. The truncation, dimensional reduction and stability analysis are then implemented in the code {\tt MSSV}, presented in Section \ref{sec:tutorial}. A first use of this code is then the verification in Section \ref{sec:construnc} that we have a consistent truncation of our 10d starting point. This is achieved thanks to a comparison of 10d and 4d equations of motion, as explained previously. Note that both codes, {\tt MSSS} and {\tt MSSV}, and both papers have compatible conventions. This could allow to use them further together, for instance {\tt MSSS} for the search of solutions and {\tt MSSV} to study the stability. We turn in Section \ref{sec:stab} to analysing the stability of the previously mentioned solution database. We start by determining and discussing in Section \ref{sec:flatdir} the generic flat directions in each of the 21 solution classes. Those appear as massless modes in the spectrum of 10d solutions. We then study the spectrum and stability of de Sitter solutions in Section \ref{sec:dS}, Minkowski solutions in Section \ref{sec:Mink} and anti-de Sitter solutions in Section \ref{sec:AdS}. We comment on the results and compare them to corresponding swampland conjectures. Finally, we summarize our findings in Section \ref{sec:conclusion} and provide an outlook.

\section{Dimensional reduction on group manifolds}\label{sec:review}

In this section, we review the details of type II flux compactifications on group manifolds in the presence of orientifolds and $D$-branes. Our starting point is the low-energy limit of type II string theories, namely 10d type II supergravities with the actions as given in \cite{Polchinski:1998rr}. We first present in Section \ref{sec:trunc} the truncation of 10d fields, commonly followed when compactifying on 6d group manifolds. We then use it in Section \ref{sec:dimred} to perform the dimensional reduction from 10d to 4d, with a focus on scalar fields and their scalar potential $V$. This reduction is implemented in the code {\tt MSSV} described in Section \ref{sec:tutorial}, and further used in Section \ref{sec:stab} to study the 4d stability of 10d compactifications.

\subsection{Truncation ansatz}\label{sec:trunc}

We present here the truncation ansatz of the 10d fields to be used to derive our 4d theory. For phenomenology, one would like to truncate to the 4d light fields, eventually providing a 4d low energy effective theory. Unfortunately, for non-Ricci flat manifolds, it is generically not known what the lightest fields are (see however recent progress in \cite{Ashmore:2019qii,Ashmore:2019rkx}). A different truncation is then usually considered on other manifolds, e.g.~those with an $\rm{SU}(3)\times\rm{SU}(3)$ structure: that truncation has been argued to correspond to a consistent truncation \cite{Gurrieri:2002wz,DAuria:2004kwe,House:2005yc,Grana:2005ny, Benmachiche:2006df, Louis:2006kb, Kashani-Poor:2006ofe, Grana:2006hr, Kashani-Poor:2007nby, Cassani:2009ck, Andriot:2018tmb}. The finite set of 4d fields kept by such a truncation contains a priori both light and heavy fields, but this set is characterised by a certain independence with respect to other fields. This has the advantage to guarantee that a solution to the 4d equations of motion is also a solution to the 10d ones, which can be useful when looking for new solutions.

In this paper we restrict ourselves to 6d manifolds being group manifolds (see reviews in \cite{Andriot:2010ju,Danielsson:2011au,Andriot:2022yyj}). Those often carry an $\rm{SU}(3)\times \rm{SU}(3)$ structure. They admit a basis of 1-forms $\{e^a\}$, $a=1,...,6$, that are left-invariant under the group action. The same holds for wedge products of $e^a$, with constant prefactors. Under a few assumptions, it was shown in \cite{Cassani:2009ck} that for compactifications on group manifolds, expanding all 10d fields in a basis of forms that are left-invariant under the group action gives rise to a consistent truncation. The resulting 4d theory is then a gauged supergravity \cite{Samtleben:2008pe, Trigiante:2016mnt}. This was proven in the absence of orientifold projections and localized sources. As detailed below, we will consider here $D$-branes and orientifolds but restrict ourselves to smeared sources. In that case, it is still expected that the 4d theory, a gauged supergravity, is a consistent truncation\footnote{\label{foot:construncOp}The argument of \cite{Cassani:2009ck} states that left-invariant modes are singlets under the group action, therefore they do not mix with other modes, providing a consistent truncation. This would a priori still apply in the presence of smeared sources, as long as the corresponding contributions are left-invariant: in particular, the internal volume forms of subspaces parallel or transverse to the sources should be given by wedge products of the $e^a$ with constant factors. This will be the case here; see \cite{Andriot:2016xvq} for a discussion of the geometric implications. We thank Davide Cassani for explaining this point to us.} and this has been explicitly checked either formally \cite{Andrianopoli:2005jv, Villadoro:2005cu, DallAgata:2009wsi, Dibitetto:2011gm} or in several examples \cite{Caviezel:2009tu, Danielsson:2011au, Petrini:2013ika}. In Section \ref{sec:construnc}, we will verify in detail that all compactifications considered in this paper give rise to a 4d theory that is a consistent truncation.

Since the left-invariant fields, to be considered here in our truncation, are not guaranteed to be the lightest fields in the theory, we will only obtain an upper bound on the smallest masses. However, for some simple group manifolds, namely nilmanifolds, recent progress in the understanding of the lightest modes \cite{Andriot:2016rdd, Andriot:2018tmb} indicate that the consistent truncation actually contains the lightest fields in the theory. However, as pointed out in \cite{Andriot:2018wzk}, it is also possible that for other group manifolds, the left-invariant fields have masses that are larger than the Kaluza--Klein scale. We refer to \cite[Sec. 5]{Andriot:2018tmb} and \cite{Andriot:2022yyj} for further related discussions.\\

On 6d group manifolds, the set of 1-forms $\{e^a\}$ satisfy the Maurer-Cartan equation
\be\label{eq:metricflux}
\d e^a = -\frac12 {f^a}_{bc} \ e^b \w e^c \,,\qquad a,b,c=1,2,...,6\,,
\ee
where the metric fluxes ${f^a}_{bc}$ are the structure constants of the Lie algebra associated to the group. A necessary condition to ensure compactness of the group manifold is to require $\sum_a f^a{}_{ab} =0$. Here, we require in addition $f^a{}_{ab} =0$ {\sl without the sum on $a$}: this amounts to choosing a certain basis for $\{e^a\}$, but also restricting to algebras that allow such bases. This choice is due to a preference in order to find a lattice and ensure compactness \cite{Andriot:2010ju, Andriot:2022yyj}.

As mentioned above, the forms that are left-invariant under the group action are wedge products of the $e^a$ with constant prefactors. The 4d scalar fields are obtained by expanding the 10d fields in terms of left-invariant forms. In this case the prefactors, i.e. the 4d scalar fields, are still functions of the 4d spacetime coordinates $x^\mu$ but they are constant as functions of the internal group manifold coordinates $y^m$. For example, the 10d dilaton gives rise in this truncation to a single real scalar $\phi_{10d}(x^\mu,y^m) \rightarrow \phi_{4d}(x^\mu) \cdot 1$, where $1$ is the left-invariant 0-form on the internal group manifold. Another example are the axions, such as $B_2 = \tfrac12 b_{ab}(x^\mu) e^a \w e^b + ...$ where the $b_{ab}(x^\mu)= -b_{ba}(x^\mu)$ are a set of 4d scalar fields that arise from the truncation and reduction of the field $B_2$. There will be additional terms in the expansion of $B_2$ that will give rise to the internal $H_3$-flux via
\ba
H_3 = \d B_2 &=& \d x^\nu \w \partial_\nu B_2 +e^a \w \partial_{a} B_2 \label{HB}\\
&=& \frac12 \partial_\nu b_{ab}(x^\mu) \d x^\nu \w e^a \w e^b -\frac12 b_{ab}(x^\mu) {f^a}_{cf} e^b \w e^c \w e^f+\frac{1}{3!} h_{abc} e^a \w e^b \w e^c \,,\nn
\ea
where $e^a= e^a{}_m(y)\, \d y^m$ and $\del_a= e^m{}_a(y)\, \del_m$. Here, $h_{abc}$ is fully antisymmetric and denotes the constant flux number threading internal 3-cycles, that should be quantized in string theory provided $e^a \w e^b \w e^c$ is a harmonic form. The same type of flux quanta will appear as constant prefactors for RR fluxes. In the above expansion we have neglected potential 4d 1-forms that would arise via $B_2 \supset A_{\mu,\,a} \d x^u \w e^a$ since we are only interested in scalar fields and we will neglect potential gauge fields in the models discussed below. We have also neglected a 4d 2-form that arises from $B_2 \supset \tfrac12 b_{\mu \nu} \d x^\mu \w \d x^\nu$ since it is projected out by orientifold projections in our settings, as discussed in the next paragraph. However, for the 10d RR forms $C_p$ there can be such 2-forms in 4d and they will have to be dualized into scalar fields using the 4d Hodge star.

Finally, for the 10d metric we assume a block diagonal unwarped form, consistently with the smeared sources we consider as well as the dilaton being independent of $y^m$
\be
ds^2 = G_{MN}\, \d x^M \d x^N = G_{\mu\nu} \d x^\mu \d x^\nu + G_{ab} e^a e^b\,.
\ee
The background 4d metric is that of a maximally symmetric spacetime: we restrict indeed to compactifications to 4d de Sitter, Minkowski or anti-de Sitter. This restricts the possible 4d background fluxes and the allowed sources. The background 6d metric should be $\delta_{ab}$; the left-invariant components $G_{ab}$ can be viewed as fluctuations around the former. Those are the 4d scalar fields generalizing the usual K\"ahler and complex structure moduli beyond Calabi-Yau compactifications.

The sources considered are $D_p$-branes and orientifold $O_p$-planes. Those are gathered in sets $\{I\}$ of parallel sources of same dimensionality $p$, i.e.~those $O_p/D_p$ wrapping the same internal directions along wedge products of the $e^a$ (see footnote \ref{foot:construncOp}). All sources are taken space-filling in 4d because of the maximal symmetry. Their appearance in the equations will be through the smeared contribution $T_{10}^{(p)_I}$ of each set $I$, defined below. We will additionally perform orientifold projections that project out part of the fields. For those, we follow the conventions of \cite[Sec. 3.1]{Koerber:2007hd}. We will mod out by the worldsheet parity operator $\Omega_p$ and a spacetime involution $\sigma$. Additionally, we will sometimes include a factor of $(-1)^{F_L}$. We will not do separate orbifold projections but rather impose multiple orientifold projections that often can be combined to find a simple orbifold projection. However, in some instance if we introduce orientifolds with different dimensionalities, like $O_4/O_6$ or $O_5/O_7$, there is a residual factor of $(-1)^{F_L}$ that acts together with what would usually be the orbifold action. Thus, it is easier to mod out by several orientifold projections.\\

Let us present an example of the set of 4d scalar fields resulting from this truncation performed together with the orientifold projections. We consider the solution class $s_{55}$: this type IIB setting includes $O_5$ (and possible $D_5$) along internal directions 12, $O_5$ $(D_5)$ along 34 and possible $D_5$ along 56. As a result, the list of 4d real fields, that are all functions of $x^{\mu}$, is given by
\begin{equation}
\begin{aligned}
       s_{55}: \quad    &G_{11} \,, G_{12}  \,, G_{22} \,, G_{33} \,, G_{34} \,, G_{44} \,, G_{55} \,, G_{56} \,, G_{66}\,, \\
       &C_{2 \ 12} \,, C_{2 \ 34}\,, C_{2 \ 56}\,, C_{4 \ 1356}\,, C_{4 \ 1456}\,, C_{4 \ 2356}\,, C_{4 \ 2456}\,, C_{6 \ 123456}\,, \\
       &b_{13}\,, b_{14}\,, b_{23}\,, b_{24}\,, \phi \,.
\end{aligned}\label{s55fields}
\end{equation}
These fields are worked out automatically in the code {\tt MSSV} presented in Section \ref{sec:tutorial}. The number of fields for each solution class to be considered is listed in Table \ref{tab:numberfields}, while we give for completeness in Table \ref{tab:numberfieldsother} the number of the fields for the other 13 solution classes of \cite{Andriot:2022way}.
\begin{table}[H]
  \begin{center}
    \begin{tabular}{|c||c|c|c|c|c|c|c|c|}
    \hline
Class & $s_{55}$ & $s_{555}$ & $s_{66}$ & $s_{6666}$ & $m_{46}$ & $m_{466}$ & $m_{55}$ & $m_{5577}$ \\
    \hhline{=::========}
$\#$ of fields & 22 & 14 & 22 & 14 & 22 & 14 & 22 & 14 \\
    \hline
    \end{tabular}
     \caption{The number of scalar fields for each solution class where new solutions have been found in \cite{Andriot:2022way}.}\label{tab:numberfields}
  \end{center}
\end{table}
In Table \ref{tab:numberfields}, the matching of the number of fields in different solution classes is remarkable. It may be understood by T-duality, as we now explain. The 10d theories are known to be generally T-dual to each other, and so should be the generic development in left-invariant fields. What matters then are the configurations of orientifolds which project out certain fields. The classes with 22 scalar fields have 2 sets with $O_p$-planes, and those sets are T-dual to each other when going from one class to the other \cite[(4.1)]{Andriot:2022way} (provided the right background isometries are there). Since the orientifold projections are crucial in fixing the number of scalar fields, it makes sense to get the same number of fields. Similarly, in the classes with 14 scalar fields, the sets with $O_p$ are T-dual to each other when going from one class to the other. There are two ways to see this. First, one can consider only 3 sets with $O_p$ there, since the fourth one in $s_{6666}$ and $m_{5577}$ is shown to bring no further projection \cite{Andriot:2022way}. Alternatively, one can add without constraint an $O_9$ to $s_{555}$ and an $O_8$ to $m_{466}$ (transverse to direction 1 in conventions of \cite{Andriot:2022way}), making the resulting configurations of $O_p$ planes T-dual to each other among the 4 classes with 14 scalars. Note that we restrict here to geometric setups: this means we allow in the NSNS sector for the $H_3$-flux and the metric fluxes ${f^a}_{bc}$ but no non-geometric fluxes. Since the metric fluxes can become non-geometric fluxes under T-duality, each of the different classes above can give rise to different 4d theories and deserves to be studied in its own right. Non-T-dual de Sitter solutions were in particular found in several of them. This observation does not change the number of fields.

\begin{table}[H]
  \begin{center}
    \begin{tabular}{|c||c|c|c|c|c|c|c|c|c|c|c|c|c|}
    \hline
Class & $s_{3}$ & $s_{4}$ & $s_{5}$ & $s_{6}$ & $s_{7}$ & $s_{77}$ & $m_{4}$ & $m_{6}$ & $m_{66}$ & $m_{5}$ & $m_{57}$ & $m_{7}$ & $m_{77}$ \\
    \hhline{=::=============}
$\#$ of fields & 38 & 38 & 38 & 38 & 38 & 22 & 38 & 38 & 22 & 38 & 22 & 38 & 22 \\
    \hline
    \end{tabular}
     \caption{The number of scalar fields for each of the remaining 13 solution classes of \cite{Andriot:2022way}, given for completeness.}\label{tab:numberfieldsother}
  \end{center}
\end{table}

\subsection{Dimensional reduction}\label{sec:dimred}

Having presented the truncation ansatz of our 10d fields, we are now ready to perform the dimensional reduction to 4d, eventually obtaining the corresponding 4d theory. As a starting point, the 10d actions for type IIA and type IIB supergravity are given in equations (12.1.10), (12.1.24) and (12.1.26) in \cite{Polchinski:1998rr}. Up to few differences to be specified, they read as follows
\ba
S^{\rm IIA/B} &=& S_{\rm NS} + S_{\rm R} + S_{\rm CS}\,,\cr
S_{\rm NS} &=&\frac{1}{2\kappa_{10}^2} \int d^{10}x \sqrt{-G} e^{-2\phi} \lp R_{10}+4 \partial_\mu \phi \partial^\mu \phi -\frac12|H_3|^2 \rp \,,\label{eq:IIaction}
\ea
with, in type IIA
\ba
S_{\rm R}&=& - \frac{1}{4\kappa_{10}^2} \int d^{10}x \sqrt{-G}\, \bigg( F_0^2+|\tilde{F}_2|^2 +|\tilde{F}_4 |^2 \bigg) \,,\cr
S_{\rm CS} &=& -\frac{1}{4\kappa_{10}^2} \int \lp 2 C_3 \w H_3 + B_2 \w \tilde{F}_4 \rp \w \tilde{F}_4 \,,\label{eq:IIAaction}
\ea
and in type IIB
\ba
S_{\rm R}&=& - \frac{1}{4\kappa_{10}^2} \int d^{10}x \sqrt{-G}\, \bigg( |F_1|^2+|\tilde{F}_3|^2 +\frac12 | \tilde{F}_5 |^2 \bigg) \,,\cr
S_{\rm CS} &=& +\frac{1}{4\kappa_{10}^2} \int F_5 \w \tilde{F}_5 \,.\label{eq:IIBaction}
\ea
In both theories, $H_3$ and $F_q$ can a priori have purely 4d components. We restrict however to maximally symmetric 4d solutions so that for example for $H_3$ only kinetic terms for a 4d 2-form can arise but for $F_q$ also spacetime filling fluxes are possible if $q \geq4$. We are interested in the scalar potential and therefore dualise 4d 2-forms to scalars and neglect vector fields and non-dynamical 3-form fields (see below \eqref{HB}). Beyond such 4d components, we use $H_3 =\d B_2$ and $F_q =\d C_{q-1}$ as in \eqref{HB}; those can include  internal {\sl background flux}. The definition of the $\tilde{F}_q$ entering the action is made delicate in presence of such background fluxes, as discussed e.g.~in \cite{DeWolfe:2005uu} and \cite[App. B]{Cassani:2008rb}, going beyond the standard textbook material. We define here
\bea
& \tilde{F}_2= F_2 + F_0\ B_2 \ ,\nn\\
& \tilde{F}_3= F_3-C_0 \ H_3 + F_1 \w B_2 \ ,\nn\\
& \tilde{F}_4= F_4+C_1 \w H_3 +F_2 \w B_2 +\tfrac{1}{2}F_0\ B_2 \w B_2 \ ,\nn\\
& \tilde{F}_5= F_5  - C_2 \w H_3 + F_3 \w B_2 - C_0 \w H_3 \w B_2 +\tfrac{1}{2} F_1 \w B_2 \w B_2 \ , \label{exprtildeF}
\eea
where in addition to 4d flux components in $F_4$ and $F_5$, fluxes $H_3$ and $F_q$ are defined as in \eqref{HB}, and $B_2, \, C_{q'}$ are now purely 4d axions (meaning their components, on the left-invariant basis of forms, are only 4d dependent axions and do not contain contributions to internal fluxes anymore).

Let us make a few remarks to help understand the above. In $S_{\rm CS}$ in type IIA\footnote{The factor of 2 in $S_{\rm CS}$ in type IIA can be understood as compensating the fact that $\tilde{F}_4$ appears twice in the other term.} as well as in expressions \eqref{exprtildeF}, first note that some terms are paired and seem related up to a total derivative: $F_q \w B_2 \pm C_{q-1} \w H_3$. It is more common to have only one of these terms appearing, as the second one could often be recovered by integration by parts. As a consequence, one would tend to have a $\tfrac{1}{2}$ in front of such a pair of terms. Here, this factor $\tfrac{1}{2}$ is morally absorbed by the split into background fluxes in $F_q,\, H_3$, and 4d axion components in $B_2, \, C_{q-1}$, which lifts the degeneracy between the terms. Pursuing the comparison to textbook material \cite{Polchinski:1998rr}, we note here the additional term $F_1 \w B_2 \w B_2$, and the presence of $C_1 \w H_3$ in $S_{\rm CS}$. There have been in the literature arguments presented for those additional terms. To start with, T-duality maps the $F_0\ B_2 \w B_2$ term in type IIA into $F_1 \w B_2 \w B_2$ which we included in the square above, as in \cite{Bergshoeff:1996ui}. This term is important and has been argued for using T-duality in the context of axion monodromy inflation in \cite{McAllister:2014mpa}. We also added this term in $S_{\rm CS}$ in type IIB, completing the whole contribution into $\tilde{F}_5$.\footnote{$S_{\rm CS}$ in type IIB could be rewritten as $F_5 \w F_5 = 0$.} Likewise, we added $C_1 \w H_3$ or $F_2 \w B_2$ in $S_{\rm CS}$ in type IIA, completing the expression of $\tilde{F}_4$. This was discussed for example in \cite[App. B]{Cassani:2008rb}, from which we also took the sign of this term; this sign change is standard in the type IIA flux compactification literature and without it the 10d solutions would not solve the 4d equations of motions. All these additional terms will be taken into account below, and will play an important role when considering the spacetime filling $F_4$ or $F_5$ flux and considering their dual, in particular $F_6$. Eventually, the resulting 4d scalar potential will match well-known results, in particular the 10d equations of motion (see Section \ref{sec:construnc}).

We will expand all fields as described above and integrate over the internal six dimensions. We also rescale the 4d metric $G_{\mu\nu}$, of determinant $G_4$ and curvature $\overline{R}_4$, towards $g_{\mu\nu}$ of determinant $g$ and curvature $R_4$, as follows: $G_{\mu\nu} = \frac{e^{2\phi}}{vol_6}\, g_{\mu\nu}$. The dimensionless internal volume of the group manifold is given by $vol_6 = (2\pi \sqrt{\alpha'})^{-6} \int d^6y \sqrt{G_6}$, where we recall the string length definition $l_s =\sqrt{\alpha'}$. This leads to the 4d Einstein frame
\ba
\int d^{10}x \sqrt{-G}\, e^{-2\phi}\, R_{10} &=& (2\pi \sqrt{\alpha'})^{6}\int d^4x \sqrt{-G_4}\, vol_6\, e^{-2\phi}\, \overline{R}_4 + ... \cr
&=& (2\pi \sqrt{\alpha'})^{6} \int d^4x \sqrt{-g}\, R_4 + ...
\ea
Using that $2\kappa_{10}^2 = (2\pi)^7 (\alpha')^4$ we can identify $M_p = \lp \pi \alpha' \rp^{-\frac12}$.\footnote{Note that different conventions would lead to $M_p$ depending on the vev of the internal volume and that of the dilaton (or the string coupling), the fields being then only fluctuations: see e.g.~\cite[(4.3)]{Andriot:2022xjh} in arbitrary dimensions. Here, the fields $\phi$ and $vol_6$ are not fluctuations but contain the full values.}
With this we obtain the 4d action in terms of the real 4d scalar fields $\varphi^i$ and the 4d Einstein frame metric $g_{\mu\nu}$
\be\label{eq:4daction}
S_{4d} = \int d^4x \sqrt{-g}\lp \frac{M_p^2}{2} R_4 -\frac12 K_{ij}\partial_\mu \varphi^i \partial^\mu \varphi^j - V(\varphi^i)\rp\,.
\ee
Since we are only interested in the 4d scalar potential $V$, we neglected in $S_{4d}$ the 4d vector fields and 3-form fields. We also dualized 4d 2-forms $b_2^i = \frac12 b_{2\, \mu\nu}^i \d x^\mu \w \d x^\nu$ into scalar fields using $\d \varphi^i=\star_4 \d b_2^i$.

Flux and gauge potential contributions to the scalar potential require some attention. With our truncation ansatz, most of them are purely internal; we recall in particular that the $B_2$ and $H_3$ 4d components are projected out by orientifolds. However, $F_4$ in type IIA and $F_5$ in type IIB can still have 4d spacetime-filling components: we denote the corresponding forms, proportional to the 4d volume form, by $F_4^{(4)}$ and $F_5^{(4)}$. To treat those properly, let us first focus on type IIA. In the action \eqref{eq:IIAaction}, the square involving $F_4$ splits into a square on internal forms and $|F_4^{(4)}|^2$. The Chern Simon's term is an integral on a 10d form, and the 4d volume form can only be found in $F_4^{(4)}$, which simplifies this term. Overall these two contributions to the action can be rewritten as
\bea
& S_{F_4^{(4)}} =  -\frac{1}{2\kappa_{10}^2}  \Bigg( \int_4 \frac12\, F_4^{(4)}\w \star_4^G\, F_4^{(4)} \, \int_6 \d^{6}y \sqrt{G_6} \\
& +\int_4 F_4^{(4)} \int_6 \Big(C_3 \w H_3 + F_4 \w B_2  + C_1 \w H_3 \w B_2 + \frac12 F_2 \w B_2  \w B_2  + \frac{1}{6} F_0\ B_2 \w B_2 \w B_2 \Big) \Bigg), \nn
\eea
where $F_4$ now denotes only internal components as in \eqref{HB}, and where the square $|F_4^{(4)}|^2$ and the Hodge star $\star_4^G$ involve $G_{\mu\nu}$. Turning to $g_{\mu\nu}$ and introducing $vol_6$, we rewrite the above as
\bea
& S_{F_4^{(4)}} =  -\frac{M_p^2}{2} \Bigg( \int_4 \frac12\, F_4^{(4)}\w \star_4^g\, F_4^{(4)} \  \frac{(vol_6)^3}{e^{4\phi}} +\int_4 F_4^{(4)} \ \, (2\pi \sqrt{\alpha'})^{-6}\! \int_6 \Big(C_3 \w H_3  \\
& \qquad\qquad\qquad\qquad+ F_4 \w B_2 + C_1 \w H_3 \w B_2 + \frac12 F_2 \w B_2  \w B_2  + \frac{1}{6} F_0\ B_2 \w B_2 \w B_2 \Big) \Bigg)\,. \nn
\eea
The last term $(2\pi \sqrt{\alpha'})^{-6}\, \int_6 ( F_4 \w B_2  + ...)$ is a dimensionless number. From there, we can apply the procedure to treat spacetime-filling fluxes: it requires to add an extra action term, and to integrate out a field. We refer to \cite[App. E.2]{Louis:2002ny}, or \cite[App. A]{Andriot:2020lea} and  \cite{Andriot:2022xjh}. We can follow here \cite[(2.15)]{Louis:2002ny} (with $B_2=0$) to replace the above action by the following one
\bea
& S_{F_4^{(4)}} \rightarrow  -\frac{M_p^2}{2} \int_4 \d^4x \sqrt{-g} \  \frac12\, \frac{e^{4\phi}}{(vol_6)^3}  \Bigg( (2\pi \sqrt{\alpha'})^{-6} \int_6 \Big( F_6 +C_3 \w H_3 + F_4 \w B_2  \\
& \qquad\qquad\qquad\qquad\qquad \quad+  C_1 \w H_3 \w B_2 + \frac12 F_2 \w B_2  \w B_2  + \frac{1}{6} F_0\ B_2 \w B_2 \w B_2 \Big) \Bigg)^2 \,. \nn
\eea
The flux $F_6$ is the internal dual to $F_4^{(4)}$ \cite{DeWolfe:2005uu, Cassani:2008rb}, i.e.~$\star_6 F_6$ captures the internal freedom of $F_4^{(4)}$. Since the dimensionless number is given in terms of the 6d integral of a 6-form, we can rewrite the above as
\beq
-\frac{M_p^2}{2}  \int_4 \d^4x \sqrt{-g} \frac12 \frac{e^{4\phi}}{vol_6} \Bigg| F_6 +C_3 \w H_3 + F_4 \w B_2  +  C_1 \w H_3 \w B_2 + \frac12 F_2 \w B_2  \w B_2  + \frac{1}{6} F_0\ B_2 \w B_2 \w B_2 \Bigg|^2 ,
\eeq
where the square involves the 6d metric. This will provide us with a corresponding term in the scalar potential below in equation \eqref{VIIA}. In type IIB, we proceed similarly. We use there the anti-self-duality of the 10d $F_5$ to get the internal freedom of $F_5^{(4)}$. One eventually generates an internal square $|F_5 + ... |^2$ that was already present in \eqref{eq:IIBaction} with internal fluxes and gauge potentials, hence effectively removing the factor $1/2$ in $S_{{\rm R}}$.

Lastly, we want to include $D_p$-brane sources and $O_p$-planes that fill the space in 4d and wrap internal $(p-3)$-dimensional spaces $\Sigma_{p-3}$ in the group manifold. Thus, we need necessarily $p\geq 3$ in order to have a maximally symmetric 4d spacetime. We will work in a smeared limit in which we do not keep track of the position of the localized sources in the internal space but we rather ``smear'' them over the internal space. We do not include the worldvolume fields for the $D_p$-branes, so that the source contributions to the 10d action are given by
\be
\hspace{-0.1in} S_{O_p/D_p} = - T_{O_p/D_p} \int \left.\lp  d^4x \, d^{\,p-3}y \ \sqrt{|G+B_2|}\, e^{-\phi} - e^{-B_2} \sum_q C_q \rp\right|_{M^{3,1}\times \Sigma_{p-3}} \hspace{-0.3in} \wedge j(\tilde{\Sigma}_{9-p})\,,
\ee
where $|G+B_2|$ denotes the absolute value of the determinant of the tensor $G+B_2$, here further pulled back to the worldvolume. The $(9-p)$-form $j(\tilde{\Sigma}_{9-p})$ can be understood as the constant unit volume form on the $(9-p)$-dimensional space $\tilde{\Sigma}_{9-p}$ that is dual on the internal manifold to $\Sigma_{p-3}$: it satisfies $\int_{\tilde{\Sigma}_{9-p}} j(\tilde{\Sigma}_{9-p})=1$. We also indicated that the fields need to be pulled back to the source worldvolume which is given by $M^{3,1}\times \Sigma_{p-3}$. Note that the pullback of $B_2$ to the worldvolume of the $D_p$-branes and $O_p$-planes vanishes for all solution classes to be considered, namely those of Table \ref{tab:numberfields}, except for $m_{55}$ and $s_{66}$.\footnote{Since $B_2$ is odd under the orientifold involution its pullback to an $O_p$-plane worldvolume is zero. However, for some $D_p$-branes the pullback of $B_2$ to the worldvolume is non-zero in our solution classes $m_{55}$ and $s_{66}$. Their contribution to the scalar potential is however quadratic, so they do not contribute to the gradient of the potential, but only to the mass matrix. On the 10d side, contributions from the source action to the 10d $B$-field equation of motion have appeared in \cite[(5.3)]{Koerber:2007hd}, but we ignored them in \cite{Andriot:2022way} where we looked for new 10d solutions; it is a priori unclear to us whether there would be such contributions for the various solution classes. Since however the contributions in $m_{55}$ and $s_{66}$ do not alter the gradient of the potential, the 10d $B$-field equation of motion is unlikely to change. In any case, our solutions then remain critical points of the potential, and the truncation is consistent with our 10d $B$-field equation.} The tension of $N_{O_p}$ $O_p$-planes is $T_{O_p} = - 2^{p-5} N_{O_p}(2\pi)^{-p} \alpha'^{\,-\frac{p+1}{2}}$, and for $N_{D_p}$ $D_p$-branes one has $T_{D_p} = N_{D_p}(2\pi)^{-p} \alpha'^{\, -\frac{p+1}{2}}$.
Note that one can in principle add an arbitrary number of $D_p$-branes but the number of $O_p$-planes is fixed by the number of fixed points of the corresponding $O_p$-plane involution. The second term in the above action $S_{Op/Dp}$ does not contribute to the 4d scalar potential, but is relevant for the (sourced) Bianchi identities and the tadpole cancellation conditions that need to be imposed in addition to the 4d equations of motion. These extra conditions can for instance be found in \cite[(2.7)]{Andriot:2016xvq} in our conventions.

The first term in the action above does contribute to the scalar potential and can be rewritten as
\ba
S_{O_p/D_p} &=& - T_{O_p/D_p} \int \left. d^4x \, d^{\,p-3}y \ \sqrt{|G+B_2|}\, e^{-\phi}\right|_{M^{3,1}\times \Sigma_{p-3}} \wedge j(\tilde{\Sigma}_{9-p})\cr
&=& - T_{O_p/D_p}\, (2\pi \sqrt{\alpha'})^{p-3} \int d^4x \sqrt{-g}\, e^{3\phi} \, \frac{ volB_{\,p-3}}{(vol_6)^2} \,,
\ea
where we introduce the following notations
\bea
& volB_{\,p-3} = (2\pi \sqrt{\alpha'})^{3-p} \int d^{\, p-3}y \left.\sqrt{|G_6+B_2|}\right|_{\Sigma_{p-3}} \ ,\\
& vol_{p-3} = (2\pi \sqrt{\alpha'})^{3-p} \int d^{\, p-3}y \left.\sqrt{|G_6|}\right|_{\Sigma_{p-3}} \ ,\nn
\eea
with $G_6$ standing for the internal components of the 10d metric, and $vol_{p-3}$ denotes the dimensionless volume of the internal space $\Sigma_{p-3}$ wrapped by the source. For a given dimensionality $p$, the sources can wrap different internal $(p-3)$-dimensional spaces $\Sigma_{p-3}^I$, $I=1,2, ...$ We recall the notion of a set $I$ of sources being along the same dimensions, and we introduce the corresponding numbers of $N_{O_p}^I$ $O_p$-planes and $N_{D_p}^I$ $D_p$-branes wrapping the same $\Sigma_{p-3}^I$. We then follow the conventions of \cite{Andriot:2016xvq,Andriot:2017jhf,Andriot:2019wrs} and use the notation $T_{10}^{(p)_I}$,\footnote{Strictly speaking, $T_{10}^{(p)_I}$ has been defined beyond the smeared case, but for $B_2|_{\Sigma_{p-3}}=0$; we naturally extend here the definition. Note that the on-shell value of this quantity is not modified, since at our critical points, axion vevs vanish.} defined in the smeared case via the following equation
\ba
\frac{M_p^2 }{2} \frac{T_{10}^{(p)_I}}{p+1} &=& \frac{1}{(2\pi)^p {\alpha'}^{\frac{p+1}{2}}}\ (2\pi \sqrt{\alpha'})^{p-3} \, (2^{p-5} N_{O_p}^I - N_{D_p}^I) \ \frac{volB_{\,p-3}^{\,I}}{vol_6}\cr
&=&  \frac{1}{(2\pi)^3 (\alpha')^2} \, (2^{p-5} N_{O_p}^I - N_{D_p}^I) \  \frac{volB_{\,p-3}^{\,I}}{vol_6} \,.
\ea
Further, we define
\be
T_{10}^{(p)} = \sum_I T_{10}^{(p)_I} \, .
\ee

Combining all the contributions above, the scalar potential is then given by
\begin{empheq}[box=\boxed]{align}
\label{VIIA}
&\ V^{\rm IIA}(\varphi^i) = \frac{M_p^2 }{2} \frac{e^{2\phi}}{vol_6} \bigg( -R_6 +\frac12 |H_3|^2_{\rm int}  - e^{\phi} \sum_{p=4,6,8}\, \frac{T_{10}^{(p)}}{p+1}  \\
& \qquad \qquad  \quad+ \frac{e^{2\phi}}{2} \bigg[ F_0^2 +\big|F_2 + F_0 B_2\big|^2_{\rm int} +\left|F_4  + C_1 \w H_3 +F_2 \w B_2  +\frac{1}{2}F_0\ B_2 \w B_2\right|^2_{\rm int} \nn\\
&  +\left|F_6 + C_3 \w H_3 + F_4 \w B_2 + C_1 \w H_3 \w B_2 +\frac{1}{2}F_2\w B_2 \w B_2 + \frac{1}{6}F_0\ B_2 \w B_2\w B_2\right|^2_{\rm int} \bigg] \bigg) \nn
\end{empheq}
\vspace{-1.1in}

\noindent in type IIA and by
\begin{empheq}[box=\boxed]{align}
\label{VIIB}
V^{\rm IIB}(\varphi^i) & = \frac{M_p^2 }{2} \frac{e^{2\phi}}{vol_6} \bigg( -R_6 +\frac12 |H_3|^2_{\rm int} - e^{\phi}\sum_{p=3,5,7,9}\, \frac{T_{10}^{(p)}}{p+1} \nn\\
& \qquad \qquad \qquad + \frac{e^{2\phi}}{2} \bigg[ |F_1|^2_{\rm int} +\big|F_3 - C_0\wedge H_3 + F_1 \w B_2\big|^2_{\rm int} \\
&\quad + \left|F_5 - C_2 \w H_3 + F_3 \w B_2 - C_0 \w H_3 \w B_2 +\frac12 F_1 \w B_2 \w B_2\right|^2_{\rm int} \bigg] \bigg)\nn
\end{empheq}
in type IIB.\footnote{We generically include $p=9$ sources in $V^{\rm IIB}$. However, a
tadpole constraint on them would require to cancel the $O_9$-plane charge with that of $D_9$-branes, meaning here $T^{(9)}_{10}=0$. Such sources would then not contribute to the potential, except indirectly through the $O_9$-projection. Because of $T^{(9)}_{10}=0$, $p=9$ sources have anyway not been considered in \cite{Andriot:2022way, Andriot:2022yyj}, i.e.~in the solutions to be discussed in Section \ref{sec:stab}.} We used $|\cdot|_{\rm int}$ to denote the contractions of the form only with respect to the internal metric $G_{ab}$ on the group manifold. As discussed around \eqref{exprtildeF}, let us recall that the fluxes in these potentials are given by (the internal forms in) \eqref{HB} including axion terms, and that the gauge fields $B_2$ and $C_q$ contain only 4d axions. These scalar potentials match expressions obtained from ${\cal N}=1$ supersymmetric compactifications, as for example those in \cite{Grimm:2004uq, Grimm:2004ua}. Moreover, they will be successfully tested against 10d equations in Section \ref{sec:construnc}.

For 6d group manifolds the Ricci scalar is given in terms of the metric fluxes defined in equation \eqref{eq:metricflux}. If one restricts to algebras for which $\sum_a {f^a}_{ab}=0$, then it is explicitly given by
\be
R_6 = -\frac12 {f^a}_{bc} {f^b}_{ae}\, G^{ce} -\frac14 {f^a}_{bc} {f^e}_{fg}\, G_{ae} {G}^{bf} {G}^{cg}\,.
\ee
This provides its contribution to the scalar potential.\\

Let us now say a word on the kinetic terms and the field space metric $K_{ij}$. Its computation is automatized in {\tt MSSV}. The kinetic terms for the axions arise from the squares of the field strengths. For example, for $F_{p+1}$ one finds following \eqref{HB}
\ba
\label{kinCp}
S_{{\rm kin,} \, F_p} &=& - \frac{1}{4\kappa_{10}^2} \int d^{10}x\ \sqrt{-G} |F_{p+1}|^2\\
& \supset& - \frac{1}{4\kappa_{10}^2} \int d^{10}x \sqrt{-G}\  \frac{1}{p!}\partial_\mu C_{p \ a_1 a_2\,...\, a_p} G^{\mu\nu} G^{a_1 b_1}G^{a_2 b_2}...\, G^{a_p b_p} \partial_\nu C_{p \ b_1 b_2\,...\, b_p}\cr
&=& - \frac{M_p^2}{4} \int d^{4}x \sqrt{-g}\, \frac{e^{2\phi}}{p!}\partial_\mu C_{p\ a_1 a_2\,...\, a_p}\, g^{\mu\nu} G^{a_1 b_1}G^{a_2 b_2}...\, G^{a_p b_p} \partial_\nu C_{p\ b_1 b_2\,...\, b_p} \,. \nn
\ea
Given our definition of the field space metric (or kinetic matrix) $K_{ij}$ above in equation \eqref{eq:4daction}, we see that the entries corresponding to the $C_p$ axions would be $M_p^2/2$ multiplied by $e^{2\phi}$ and a combination of components of the inverse internal metric. In particular, at the special point where we set all diagonal metric entries as well as $e^{2\phi}$ to one and the off-diagonal ones to zero, all the $C_p$ axions as well as all $B_2$ axions will give rise to a diagonal $K_{ij}$ submatrix with entries $M_p^2/2$.

In order to find the kinetic terms for the dilaton and the scalars $G_{ab}(x^\mu)$ arising from the internal metric one has to calculate the Ricci scalar $R_{10}$. After doing the rescaling to 4d Einstein frame and after doing appropriate integrations by parts of second derivative terms, one can subtract the background $R_4$ and $R_6$ and then add the $4 \partial_\mu \phi \partial^\mu \phi$ term from equation \eqref{eq:IIaction} to get the final kinetic terms. An illustration of this procedure on few fields can be found in \cite[App. D]{Andriot:2020wpp}.
\\

Neglecting flux quantization, as well as quantization of $N_{O_p/D_p}^I$ and that of the structure constants \cite{Andriot:2020vlg}, the type II supergravity actions have a large symmetry group that allows one to rescale and shift the fluxes. This enables us to move any given point in field space to the point where all axionic scalar fields and all off-diagonal metric entries are equal to zero and all diagonal metric entries and the dilaton $e^\phi$ are equal to one. Since the classical scalar potential has a complicated dependence on the fields but is quadratic in the fluxes and linear in the sources, it is much easier to solve the equations of motions and find critical points at a fixed point in field space in terms of the flux parameters. One can then use the rescaling and shifts of the fluxes to move the critical point. If one can find in this way a point in field space at large volume, large complex structure and weak string coupling, where also the necessary quantization is obeyed, then one can trust the corresponding solution as a classical string background.

Given that most corrections to the classical flux potential are not known, it is not exactly clear where the trustworthy large volume and weak coupling regime begins. We therefore do not focus on this point but are rather interested in the mass spectrum for a given critical point. While the actual masses of the fields can change under the above rescalings (see e.g.~the $\lambda$-rescaling discussed in \cite[Sec. 4.2]{Andriot:2020wpp}), the $\eta_V$ parameter defined in \eqref{etaV} below cannot. This also means that the number of tachyons cannot change under this rescaling. This will be important and sufficient for our analysis below that we can hence carry out at any point in field space.

From the field space metric $K_{ij}$ defined in \eqref{eq:4daction} and discussed around \eqref{kinCp}, one obtains the mass matrix
\beq
M^i{}_k = K^{ij} \nabla_j \partial_k V \ , \label{massM}
\eeq
where $\nabla_j v_k = \del_j v_k - {\Gamma^l}_{jk} v_l$ is the covariant derivative on $v_k$ and
${\Gamma^l}_{jk}$ denotes the Christoffel symbol associated with $K_{ij}$. The eigenvalues of $M$ are the masses${}^2$. Considering its minimal eigenvalue, denoted by ``min'', one defines for $V\neq 0$ the parameter
\be
\eta_V = M_p^2\ \frac{\text{min} \lp \, K^{ij} \nabla_j \partial_k V \, \rp}{V}\,. \label{etaV}
\ee
As explained above, the sign and value of $\eta_V$ are not sensitive to rescalings of the fields, so it will be a useful parameter in the following, when studying the stability of solutions in the context of the swampland program. We also introduce for convenience the parameter
\beq
\epsilon_V = \frac{M_p^2}{2} \left( \frac{\sqrt{K^{ij} \partial_i V  \partial_j V }}{V} \right)^2 \ . \label{epsV}
\eeq

The fields $\{\varphi^i\}$ to be considered can always be brought to a canonical basis $\{\hat{\varphi}^i\}$, where the field space metric $K_{ij}$ becomes $\delta_{ij}$, i.e.~the kinetic terms become diagonal and normalized. This change of basis is given by the field space diffeomorphism $P^i{}_j=\frac{\del \hat{\varphi}^i}{\del \varphi^j}$. In terms of matrices, one has $(K_{ij}) = P^T \delta P$, where $\delta$ is simply the identity matrix; we refer to \cite[(2.2)]{Andriot:2021rdy} for more detail. The fields one-forms (or here their coefficients) are obviously related by $\del_{\mu} \hat{\varphi}^i = P^i{}_j \del_{\mu} \varphi^j $, while the vectors obey $P^i{}_j \del_{\hat{\varphi}^i} = \del_{\varphi^j}$; the latter is written in matrix form $\del_{\hat{\varphi}} \, P = \del_{\varphi}$ where the derivatives are one-line matrices. One can then consider the mass matrix $\hat{M}$ in the canonical basis, of coefficients $\hat{M}^i{}_k = \delta^{ij} \nabla_{\hat{j}} \partial_{\hat{k}} V$. One can verify that $\hat{M} = P M P^{-1}= \delta^{-1} P^{-T}\nabla \del V P^{-1}$, where the matrix $\nabla \del V$ computes the Hessian of $V$ in the non-canonical field basis. We will use these formulas to get our spectrum data. Indeed, the code {\tt MSSV} first computes the matrix $P$ using the relation $(K_{ij}) = P^T \delta P$, and introducing an orthonormal matrix $O$ and a diagonal one $D$, such that $P=D^{-1}O$. Then, the mass matrix is computed with $\hat{M} = \delta^{-1} P^{-T}\nabla \del V P^{-1}$. From there, its eigenvalues and eigenvectors are determined. An eigenvector can be expressed as $v = v^{i}\, \del_{\hat{\varphi}^i}$, and we store the coefficients $v^{i}$ as columns in a matrix. Expressing the eigenvectors in the non-canonical basis (to understand their field directions) then amounts to using the relation $\del_{\hat{\varphi}} = \del_{\varphi} \, P^{-1}$. We will obtain in this way the spectrum data, and use it to analyse the stability of solutions in Section \ref{sec:stab}.

\section{The code {\tt MSSV}: tutorial}\label{sec:tutorial}

In this section, we briefly present how the code {\tt MSSV} works, and provide a few useful commands. {\tt MSSV} stands for Maximally Symmetric spacetime Solutions $V$, where $V$ refers to the complete scalar potential $V$ obtained via the dimensional reduction described above. The code has been developed \texttt{Wolfram Mathematica 13} \cite{Mathematica}. {\tt MSSV} takes as input the source configuration corresponding to a certain solution class of \cite{Andriot:2022way}, meaning the number of $O_{p}$-planes, of $D_{p}$-branes, and the directions along which they are placed. This is also referred to as the ``model'', since this data determines the 4d theory and its field content. Once executed, the code begins by using the user's input to find the left-invariant scalar fields and the background fluxes after the orientifold projections. The code then computes, following Section \ref{sec:review}, the 4d kinetic terms and scalar potential for these fields and fluxes and packages them as per equation \eqref{eq:4daction}. The code also determines those of the scalar fields that are generic flat directions, as discussed in Section \ref{sec:flatdir}. In the code we have set $M_p=1$.

The next required input is that of a concrete 10d solution obtained from the code {\tt MSSS} and listed in the accompanying database \cite{Andriot:2022way}. This amounts to assigning values to all the fluxes appearing in the variable \texttt{fluxes}. Through the command {\tt AnalyseSol}, after performing a few checks on the 10d solution, the code is then able to extract the mass spectrum and the $\eta_V$ value (see equations \eqref{massM} and \eqref{etaV}). It also provides the mass matrix eigenvectors for the tachyonic and massless modes. In the case of multiple massless modes, the corresponding eigenspace is degenerate and the code chooses a random basis in field space that spans it.

Note that the computation of the mass spectrum by the code assumes to be at a critical point, where $\partial_{\varphi^i} V=0$, axions (including off-diagonal metric components) vanish, and diagonal metric components as well as $e^{\phi}$ are set to 1. All models analysed in Section \ref{sec:stab} do correspond to critical points. In order to calculate the kinetic terms, the potential value, $\epsilon_V$ and $\eta_{V}$ at a generic point in field space, the user has to call \texttt{AnalysePointGen}. To run this command, the user must assign values to the background fluxes as well as to the scalar fields.

We now list a few useful commands; more are provided in the code.
\begin{itemize}
    \item \texttt{RunModel} -- Initializes the model, then prints out the left-invariant scalar fields and background fluxes for the chosen source configuration along with any generic flat directions, i.e. left-invariant scalar fields that generically do not appear in $V$.

    \item \texttt{AnalyseSol} -- Computes and prints out information regarding the mass spectrum, including masses of the fields, number of massless fields and their field directions, number of tachyons and their field directions, etc. This is computed for a given solution that satisfies $\partial_{\varphi^i} V=0$ at the critical point defined in the variable \texttt{extremum}.

    \item \texttt{AnalysePointGen} -- Returns the value of the potential, the gradient, $\epsilon_V$ and $\eta_{V}$ at a generic point in field space.

    \item \texttt{fields} -- Returns the left-invariant scalar fields for the chosen model.

    \item \texttt{fluxes} -- Returns the set of $F_{p}$, $H_{3}$, and metric fluxes for the chosen source configuration. Recall that we set ${f^a}_{ab}=0$ without the summation over $a$.

    \item \texttt{VGen} -- Returns the scalar potential as a function of the left-invariant scalar fields at a generic point in field space.

    \item \texttt{V} -- Returns the scalar potential as a function of the flux parameters defined in the variable \texttt{fluxes} and evaluated at the critical point defined in the variable \texttt{extremum}.
\end{itemize}

The two most useful commands are \texttt{RunModel} and \texttt{AnalyseModel}, which must be evaluated before calling on the other commands. Finally, let us mention that the notebook allows to analyse different source configurations as well as several solutions, without having to quit or restart it. Indeed, the main part of the code is run once and for all, and one can then just call commands, or redefine the input. This allows in particular the user to evaluate once a whole notebook where many solutions have been entered to be analysed.

\section{Consistent truncations}\label{sec:construnc}

In this section, we verify explicitly that the 4d theory obtained by the truncation and dimensional reduction described in Section \ref{sec:review} is a consistent truncation of our 10d starting point for a 4d maximally symmetric spacetime. We recall that (smeared) $O_p/D_p$ sources are present in our compactification. We verify the consistent truncation for all 21 solution classes of \cite{Andriot:2022way}.

To prove the consistent truncation, it suffices to show that the 4d and 10d actions yield the same equations of motion (eoms). On the 4d side, the eoms for the scalar fields $\varphi^i$ at an extremum (corresponding to a maximally symmetric spacetime), denoted by ``ext'', read
\begin{equation}
\label{eomsc}
   \left.\frac{\partial}{\partial \varphi^i} V \right|_{\rm ext} = 0 \, .
\end{equation}
Note that we restrict ourselves here to solutions without kinetic energy. The (trace of the) Einstein equation reads, for a maximally symmetric spacetime,
\begin{equation}
\label{eomR4}
    R_4 - 4 V_{\rm ext} = 0 \,,
\end{equation}
and we use in this section $M_p=1$.

On the 10d side, the equations of motion are the flux eoms, denoted schematically $F_i = 0$, the 6d Einstein equations $E^{ab} = 0$, the 4d Einstein equation $E_4 = 0$, and the dilaton eom $D = 0$. These can be found e.g. in \cite{Andriot:2022way}. For instance, we define
\begin{equation}
E_4 =  R_4 - \sum_{p} \frac{T_{10}^{(p)}}{p+1} + \sum_{q=0}^6 |F_q|^2 \ ,\quad   D= 2 R_{4}+ 2 R_6 + \sum_p \frac{T_{10}^{(p)}}{p+1} -|H_3|^2 \ .
\end{equation}
Note that in \cite{Andriot:2022way}, the authors considered the trace-reversed 6d Einstein equations, and considered linear combinations of eoms to obtain $E_4$, thus one expects that the matching with equations \eqref{eomsc} and \eqref{eomR4} should only hold up to taking linear combinations. Let us add that the metric used in \cite{Andriot:2022way} is $\delta_{ab}$, allowing to raise 6d Einstein equation indices towards $E^{ab}$.

For each class of solutions, we determine in components the 10d eoms using the code \texttt{MSSS} \cite{Andriot:2022way}. We remove from those the equations trivially satisfied. In 4d, we compute the components of the generic gradient at the extremum, $\left.\partial_{\varphi^i} V\right|_{\rm ext}$, with the help of \texttt{MSSV}. At this stage, one can already check that the number of 6d Einstein equations matches with the number of scalar fields arising from the internal metric, and that the number of 10d flux eoms matches with the number of axions appearing in the potential (that is, without counting the fields associated to generic flat directions).
From these two lists of equations, one can then verify the matching, which goes as follows.

For the fluxes, one has (after the appropriate labelling)
\begin{equation}
  F_i = 2 \left.\frac{\partial}{\partial a^i} V \right|_{\rm ext} \,,
\end{equation}
where $a^i$ denotes the $i$-th axion. The non-diagonal 6d Einstein equations $E^{ab}$ correspond to the variations with respect to the non-diagonal metric scalar fields $G_{ab}$ simply via
\begin{equation}
   E^{ab} = \left. \frac{\partial}{\partial {G_{ab}}} V \right|_{\rm ext} \,,  \qquad a \neq b \,.
\end{equation}
The diagonal ones are related by
\begin{equation}
     2 E^{aa} +\delta^{aa} E^{bc}\delta_{bc} = 4 \left. \frac{\partial}{\partial {G_{aa}}} V \right|_{\rm ext} \,,
\end{equation}
where the above 6d trace accommodates the fact that $E^{ab}$ correspond here to the trace-reversed equations. Finally, the dilaton and the 4d Einstein equation combine as follows
\begin{equation}
    2 E_4 - D = 2 \left. \frac{\partial}{\partial {\phi}} V \right|_{\rm ext} \,, \qquad -E_4 + D = R_4 - 4 V_{\rm ext}\,.
\end{equation}

It is remarkable that the same matching of equations works for all solution classes considered, thus proving in each case the consistent truncation for maximally symmetric spacetimes. Although this is not surprising, given all other working examples in the literature recalled in Section \ref{sec:trunc}, this remains a non-trivial check. It was made possible thanks to the two codes {\tt MSSS} and {\tt MSSV} that generate all equations to be considered for all solution classes, using the same conventions. Finally, let us recall from \cite{Andriot:2022way} that finding 10d solutions would require in addition to solve the flux Bianchi identities (including the tadpole conditions) and the Jacobi identities on the $f^{a}{}_{bc}$, to guarantee having a group manifold.

\section{Stability analysis}\label{sec:stab}

Stability of solutions with maximally symmetric spacetimes is at the heart of several swampland conjectures, as recalled in the Introduction; it also plays an important role for phenomenological models. In this section, we use the 4d theory discussed in Section \ref{sec:review} and the corresponding code {\tt MSSV} described in Section \ref{sec:tutorial} to study the stability of the de Sitter, Minkowski and anti-de Sitter solutions found in \cite{Andriot:2020wpp,Andriot:2021rdy,Andriot:2022way}. The solutions database can be found in two files provided with \cite{Andriot:2022way}. These solutions were found with the code {\tt MSSS} \cite{Andriot:2022way} which is compatible with the present code {\tt MSSV}. The conventions of \cite{Andriot:2016xvq,Andriot:2017jhf} followed in \cite{Andriot:2022way} are the same as in the present paper, and we verify in particular that these solutions are critical points of the scalar potential $V$ obtained here, satisfying $\nabla V=0$ as well as $V= \tfrac{M_p^2}{4} R_4$. This is actually formally ensured thanks to the analysis described in Section \ref{sec:construnc} regarding consistent truncations. The stability of these solutions has already been analysed using {\tt MSSSp} in \cite{Andriot:2022yyj}, considering only the 4-6 scalar fields $(\rho,\tau,\sigma_I)$ corresponding to some dilaton  and diagonal metric fluctuations. Thanks to the above dimensional reduction, embedded in the code {\tt MSSV}, we now have a complete set of scalar fields, a corresponding scalar potential as well as kinetic terms. This allows us here to provide a more complete analysis of the perturbative stability of these solutions. The latter will essentially be discussed in terms of the parameter $\eta_V$ defined in \eqref{etaV}, or the spectrum of masses${}^2$, which are the eigenvalues of the mass matrix \eqref{massM}. Both are evaluated at the critical point of the potential, corresponding to the solution, where the axions and off-diagonal metric components vanish, while the diagonal ones and the exponential of the dilaton are equal to one. Note that thanks to the lemma in \cite[Sec. 3.3]{Andriot:2020wpp}, we know that adding more scalar fields, thus increasing the size of the mass matrix, can only lower its minimal eigenvalue. As a physics consequence, solutions are expected to be more unstable here than they were found to be in \cite{Andriot:2022yyj}. We also recall from Section \ref{sec:trunc} that the 4d theory used here is unlikely to be a low energy effective theory, but is rather a consistent truncation. Therefore, we analyse the stability using modes that are not necessarily the lightest, but form an independent set with respect to other, truncated modes. Since the masses${}^2$ obtained this way give upper bounds, our conclusions on instabilities should be sufficient. On the contrary, any observed stability could only be claimed among this set of fields, and not beyond.

We first discuss in Section \ref{sec:flatdir} the appearance of flat directions in the various solution classes defined in \cite{Andriot:2022way}. We then turn successively to the stability of de Sitter, Minkowski and anti-de Sitter solutions, in the respective sections \ref{sec:dS}, \ref{sec:Mink} and \ref{sec:AdS}.

\subsection{Flat directions}\label{sec:flatdir}

Compared to the partial stability analysis of \cite{Andriot:2022yyj}, a new phenomenon is here the presence of massless modes in the spectrum of all de Sitter, Minkowski or anti-de Sitter solutions in certain solution classes. This should be distinguished from the massless mode discussed in the Massless Minkowski Conjecture \cite{Andriot:2022yyj}, systematically observed to be present for Minkowski solutions among the fields $(\rho,\tau,\sigma_I)$, but not in de Sitter or anti-de Sitter solutions.

Massless modes observed in solution classes for any cosmological constant can naturally be interpreted as being flat directions. Indeed, specifying a solution class fixes the orientifolds, it thus determines a set of scalar fields and their generic potential, independently of the cosmological constant. We verify this interpretation by explicitly identifying scalar fields that do not appear in the generic potential of their solution class. We list those in Table \ref{tab:flatdirections} for the solution classes to be considered in the subsequent stability analysis, and in Table \ref{tab:flatdirectionsother} for the remaining 13 classes of \cite{Andriot:2022way}.
\begin{table}[H]
  \begin{center}
    \begin{tabular}{|c||c|c|c|c|c|c|c|c|}
    \hline
Class & $s_{55}$ & $s_{555}$ & $s_{66}$ & $s_{6666}$ & $m_{46}$ & $m_{466}$ & $m_{55}$ & $m_{5577}$ \\
    \hhline{=::========}
Flat dir. field & $C_6$ (1) & $C_6$ (1) & $C_5$ (1) & $\varnothing$ & $C_5$ (3) & $C_5$ (1) & $C_6$ (1) & $\varnothing$ \\
    \hline
    \end{tabular}
     \caption{We consider all solution classes of \cite{Andriot:2022way} for which the stability of a solution will be analysed in Section \ref{sec:stab}. For each of them, the dimensional reduction described in Section \ref{sec:dimred} provides a finite set of fields $\{\varphi^i\}$ and a scalar potential $V$. We list in this table the fields $\varphi^i$ such that $\del_{\varphi^i} V=0$ generically: this means $\varphi^i$ is a flat direction. While we actually provide in the table a $p$-form, the number in parentheses corresponds to the amount of its components remaining after the orientifold projections, thus to the number of scalar fields being generic flat directions in the solution class.}\label{tab:flatdirections}
  \end{center}
\end{table}

The flat directions identified in Table \ref{tab:flatdirections} correspond to some RR axions. So they can indeed be distinguished from the Minkowski massless mode conjectured to be among $(\rho,\tau,\sigma_I)$. It is easy to understand why these axions do not appear in their scalar potential. In type IIB, $C_6$ could only enter the potential \eqref{VIIB} through an $F_7$-flux (the dual of a 4d $F_3$-flux): it would appear through a term proportional to a geometric flux coming from $\d C_6$. This 6d 7-form is however obviously vanishing. Therefore $C_6$ only appears in kinetic terms as a fluctuation, and is then a flat direction. $F_3$ and thus $C_6$ are however odd under an $O_7$ involution, so $C_6$ has to be projected out by an $O_7$, as in $m_{5577}$. On the contrary, in type IIA, $C_5$ can appear in the potential \eqref{VIIA} through $F_6$, the dual of a spacetime-filling $F_4$. A potential term for $C_5$ would appear through $\d C_5$, a maximal 6d form proportional to $\sum_a f^a{}_{ab}$. We however require the latter sum to vanish (implemented in our ansatz) due to the 6d compactness. So (all components of) $C_5$ are also flat directions. $C_5$ is odd under an $O_6$-plane involution: one then verifies that four $O_6$-planes as placed in $s_{6666}$ project out all $C_5$ scalar fields. The absence of these flat directions in $s_{6666}$ and $m_{5577}$, as indicated in Table \ref{tab:flatdirections}, is consistent with the observation in the next subsections that the only solutions without massless mode belong to these two classes.

\begin{table}[H]
  \begin{center}
    \begin{tabular}{|c||c|c|c|c|c|c|c|c|}
    \hline
Class & $s_{3}$ & $s_{7}$, $m_{7}$ & $s_{4}$, $m_{4}$ & $s_{5}$, $m_{5}$ & $s_{6}$, $m_{6}$ & $s_{77}$, $m_{77}$ & $m_{66}$ & $m_{57}$ \\
    \hhline{=::========}
Flat dir. field & $C_4$ (15) & $\varnothing$ & $C_5$ (5) & $C_6$ (1) & $C_5$ (3) & $C_{4 \ 3456}$ & $C_5$ (1) & $\varnothing$ \\
    \hline
    \end{tabular}
     \caption{Analogous table to Table \ref{tab:flatdirections}, indicating flat directions for the other 13 classes of \cite{Andriot:2022way}, for completeness. For $s_{77}$ and $m_{77}$, with $O_7$ placed along internal directions 1234 and 1256, only one of the three $C_4$ components is a flat direction.}\label{tab:flatdirectionsother}
  \end{center}
\end{table}
\vspace{-4mm}

To those generic flat directions in solution classes, one may add more flat directions appearing when setting to zero (generically, or even in a solution) some contribution to the scalar potential, e.g.~a background flux. Let us consider as an example $C_4$, which only appears in the potential \eqref{VIIB} through $F_5$, i.e.~as $\d C_4$ proportional to $f^a{}_{[bc} C_{4\, def]a}$. In $s_{55}$, as indicated in \eqref{s55fields}, the following $C_4$ components are possible: $C_{4\ 1356}, C_{4\ 1456}, C_{4\ 2356}, C_{4\ 2456}$. The allowed structure constants (by $O_5$ projections) that could contribute to the potential of $C_4$ are then $f^a{}_{bc}$ with $a=5,6$, $bc=13,14,23,24$. As a consequence, if one considers as a ``subclass'' of $s_{55}$ the one with these 8 $f^a{}_{bc}$ vanishing, then one gets ``generically'' in this subclass the 4 axions of $C_4$ being flat directions. The Minkowski solution $s_{55}^0 1$ can be viewed as part of such a subclass, since these 8 structure constants vanish in this solution.\\

This analysis of flat directions will help us understanding some of the massless modes appearing in the following. To conclude, let us add a remark: RR axions enjoy a continuous shift symmetry in supergravity that is broken in string theory to a discrete shift symmetry. This makes their moduli space compact and one in principle does not have to stabilize them for phenomenology. Unless one breaks this symmetry by appropriate fluxes one expects that these axions will remain flat directions at the perturbative level. Non-perturbative effects are however generically leading to a sinusoidal potential for these axions (see for example \cite[Sec. 2]{Svrcek:2006hf}). The size of these effects is model dependent and we will not study it here.

\subsection{De Sitter solutions}\label{sec:dS}

We compute for each de Sitter solution the mass spectrum with {\tt MSSV}. We report in Table \ref{tab:eta+IIA}, \ref{tab:eta+IIB1} and \ref{tab:eta+IIB2} the values of the $\eta_V$ parameter, comparing them to the values obtained with the restricted set of fields $(\rho,\tau,\sigma_I)$. We also give the number of massless modes and the number of tachyons. Those can be compared to the total number of fields in each class (Table \ref{tab:numberfields}) and the number of generic flat directions (Table \ref{tab:flatdirections}).\\

\begin{table}[ht]
  \begin{center}
  \noindent\makebox[\textwidth]{
    \begin{tabular}{|c||c|c|c|c|c|c|c|c|c|}
    \hline
class & \multicolumn{1}{c||}{$s_{66}^+$} & \multicolumn{4}{c||}{$s_{6666}^+$} & \multicolumn{4}{c|}{$m_{46}^+$}  \\
\hhline{-||-||----||----}
solution & \multicolumn{1}{c||}{1} & 1 & 2 & 3 & \multicolumn{1}{c||}{4} & 1 & 2 & 3 & 4 \\
    \hhline{=::=::====::====}
$-\eta_V$ \cite{Andriot:2022yyj} & \multicolumn{1}{c||}{3.6170} & 18.445 & 2.6435 & 2.3772 & \multicolumn{1}{c||}{3.6231} & 3.6764 & 3.7145 & 2.2769 & 2.8266 \\
    \hhline{=::=::====::====}
$-\eta_V$ & \multicolumn{1}{c||}{3.7405} & 20.836 & 2.8604 & 4.7167 & \multicolumn{1}{c||}{3.8438} & 4.0177 & 4.0679 & 3.6681 & 3.6321  \\
    \hhline{=::=::====::====}
$m^2 \leq 0$ & \multicolumn{1}{c||}{$1^-, 4^0$} & $1^-, 0^0$ & $2^-, 0^0$ & $1^-, 1^0$ & \multicolumn{1}{c||}{$1^-, 1^0$} & $2^-, 4^0$ & $1^-, 4^0$ & $2^-, 4^0$ & $2^-, 4^0$ \\
    \hline
  \multicolumn{10}{c}{}\\
\hhline{-------~~~}
class & \multicolumn{6}{c|}{$m_{46}^+$} & \multicolumn{3}{c}{} \\
\hhline{-||------|~~~}
solution  & 5 & 6 & 7 & 8 & 9 & \multicolumn{1}{c|}{10} & \multicolumn{3}{c}{} \\
    \hhline{=::======:~~~}
$-\eta_V$ \cite{Andriot:2022yyj} & 0.36462 & 3.0124 & 2.0672 & 2.3554 & 2.6418 & \multicolumn{1}{c|}{1.2539} & \multicolumn{3}{c}{} \\
    \hhline{=::======:~~~}
$-\eta_V$ & 5.1535 & 3.7518 & 3.5399 & 5.9109 & 3.8699 & \multicolumn{1}{c|}{8.1124} & \multicolumn{3}{c}{} \\
    \hhline{=::======:~~~}
$m^2 \leq 0$ & $2^-, 4^0$ & $1^-, 4^0$ & $2^-, 4^0$ & $2^-, 4^0$ & $2^-, 4^0$ & \multicolumn{1}{c|}{$1^-, 4^0$} & \multicolumn{3}{c}{} \\
\hhline{-------~~~}
    \end{tabular}
    }
     \caption{Spectrum information for each de Sitter solution in type IIA. We first provide the value of $-\eta_V$ for the fields $(\rho,\tau,\sigma_I)$ obtained in \cite{Andriot:2022yyj}, then the one obtained here with the complete set of scalar fields of the above dimensional reduction. By $i^-, j^0$, we also indicate the number $j$ of massless modes and $i$ of tachyons.}\label{tab:eta+IIA}
  \end{center}
\end{table}

\begin{table}[ht]
  \begin{center}
    \noindent\makebox[\textwidth]{
    \begin{tabular}{|c||c|c|c|c|c|c|c|c|c|}
    \hline
class & \multicolumn{9}{c|}{$s_{55}^+$} \\
\hhline{-||---------}
solution  & 1 & 2 & 3 & 4 & 5 & 6 & 7 & 8 & 9 \\
    \hhline{=::=========}
$-\eta_V$ \cite{Andriot:2020wpp} & 2.8544 & 2.7030 & 2.9334 & 2.8966 & 2.9703 & 2.9146 & 2.5101 & 2.7790 & 2.2494 \\
    \hhline{=::=========}
$-\eta_V$  & 3.9131 & 3.8971 & 3.9214 & 3.9370 & 3.9022 & 3.9063 & 3.8974 & 3.8532 & 3.9062 \\
    \hhline{=::=========}
$m^2 \leq 0$  & {$1^-, 4^0$} & {$2^-, 4^0$} & {$1^-, 4^0$} & {$1^-, 4^0$} & {$1^-, 4^0$} & {$1^-, 4^0$} & {$2^-, 4^0$} & {$1^-, 4^0$} & {$2^-, 4^0$} \\
    \hline
  \multicolumn{10}{c}{}\\
    \hline
class & \multicolumn{9}{c|}{$s_{55}^+$} \\
\hhline{-||---------}
solution  & 10 & 11 & 12 & 13 & 14 & 15 & 16 & 17 & 18 \\
    \hhline{=::=========}
$-\eta_V$ \cite{Andriot:2020wpp,Andriot:2021rdy} & 2.0908 & 2.9354 & 2.7548 & 2.9518 & 1.7067 & 2.9336 & 2.8404 & 2.8748 & $- 3.7926$ \\
    \hhline{=::=========}
$-\eta_V$  & 2.7609 & 3.9209 & 3.5411 & 3.5950 & 4.0847 & 4.2994 & 3.7656 & 3.7224 & $17.5906$ \\
    \hhline{=::=========}
$m^2 \leq 0$  & {$2^-, 4^0$} & {$1^-, 4^0$} & {$2^-, 4^0$} & {$1^-, 4^0$} & {$1^-, 4^0$} & {$1^-, 4^0$} & {$1^-, 4^0$} & {$1^-, 4^0$} & {$2^-, 4^0$} \\
    \hline
  \multicolumn{10}{c}{}\\
    \hline
class & \multicolumn{9}{c|}{$s_{55}^+$} \\
\hhline{-||---------}
solution  & 19 & 20 & 21 & 22 & 23 & 24 & 25 & 26 & 27 \\
    \hhline{=::=========}
$-\eta_V$ \cite{Andriot:2021rdy} & 0.12141 & 1.3624 & 1.7813 & 1.0525 & 1.2253 & 0.95955 & 0.90691 & 1.0438 & 1.1172 \\
    \hhline{=::=========}
$-\eta_V$  & 2.5948 & 6.4415 & 4.3007 & 2.4940 & 4.0269 & 2.7322 & 3.0085 & 3.9184 & 3.9970 \\
    \hhline{=::=========}
$m^2 \leq 0$  & {$2^-, 4^0$} & {$1^-, 4^0$} & {$1^-, 4^0$} & {$2^-, 4^0$} & {$2^-, 4^0$} & {$2^-, 4^0$} & {$2^-, 4^0$} & {$2^-, 4^0$} & {$2^-, 4^0$} \\
    \hline
    \end{tabular}
    }
     \caption{Spectrum information for each de Sitter solution in type IIB. We first provide the value of $-\eta_V$ for the fields $(\rho,\tau,\sigma_I)$ obtained in \cite{Andriot:2020wpp,Andriot:2021rdy}, then the one obtained here with the complete set of scalar fields of the above dimensional reduction. By $i^-, j^0$, we also indicate the number $j$ of massless modes and $i$ of tachyons.}\label{tab:eta+IIB1}
  \end{center}
\end{table}

\begin{table}[ht]
  \begin{center}
  \noindent\makebox[\textwidth]{
    \begin{tabular}{|c||c|c|c|c|c|c|c|c|c|}
    \hline
class & \multicolumn{1}{c||}{$s_{55}^+$} & \multicolumn{4}{c||}{$m_{55}^+$} & \multicolumn{4}{c|}{$m_{5577}^+$} \\
\hhline{-||-||----||----}
solution & \multicolumn{1}{c||}{28} & 1 & 2 & 3 & \multicolumn{1}{c||}{4} & 1 & 2 & 3 & 4 \\
    \hhline{=::=::====::====}
$-\eta_V$ \cite{Andriot:2022yyj} & \multicolumn{1}{c||}{3.2374} & 2.5435 & 2.6059 & 2.7126 & \multicolumn{1}{c||}{3.3574} & 4.7535 & 3.5034 & 3.2722 & 3.1779 \\
    \hhline{=::=::====::====}
$-\eta_V$ & \multicolumn{1}{c||}{3.7586} & 3.4316 & 3.4460 & 3.4221 & \multicolumn{1}{c||}{3.8729} & 32.725 & 3.7931 & 3.8289 & 3.7733 \\
    \hhline{=::=::====::====}
$m^2 \leq 0$ & \multicolumn{1}{c||}{$2^-, 4^0$} & $2^-, 3^0$ & $1^-, 3^0$ & $2^-, 3^0$ & \multicolumn{1}{c||}{$1^-, 3^0$} & $2^-, 0^0$ & $2^-, 0^0$ & $1^-, 1^0$ & $1^-, 1^0$ \\

    \hline
  \multicolumn{10}{c}{}\\
    \hline
class & \multicolumn{8}{c||}{$m_{5577}^+$} & \multicolumn{1}{c|}{$m_{5577}^{*\,+}$} \\
\hhline{-||--------||-}
solution  & 5 & 6 & 7 & 8 & 9 & 10 & 11 & \multicolumn{1}{c||}{12} & \multicolumn{1}{c|}{1} \\
    \hhline{=::========::=}
$-\eta_V$ \cite{Andriot:2022yyj} & 4.7957 & 4.9129 & 3.4210 & 3.5611 & 2.9333 & 2.9003 & 3.4806 & \multicolumn{1}{c||}{2.8966} & \multicolumn{1}{c|}{5.0483} \\
    \hhline{=::========::=}
$-\eta_V$  & 5.0140 & 5.1358 & 3.7551 & 3.9213 & 3.7903 & 3.8044 & 3.9849 & \multicolumn{1}{c||}{3.4820} & \multicolumn{1}{c|}{5.2673} \\
    \hhline{=::========::=}
$m^2 \leq 0$  & $1^-, 1^0$ & $1^-, 1^0$ & $2^-, 0^0$ & $1^-, 0^0$ & $2^-, 0^0$ & $2^-, 0^0$ & $1^-, 0^0$ & \multicolumn{1}{c||}{$2^-, 0^0$} & \multicolumn{1}{c|}{$1^-, 1^0$} \\
    \hline
    \end{tabular}
    }
     \caption{Spectrum information for each de Sitter solution in type IIB. We first provide the value of $-\eta_V$ for the fields $(\rho,\tau,\sigma_I)$ obtained in \cite{Andriot:2022yyj}, then the one obtained here with the complete set of scalar fields of the above dimensional reduction. By $i^-, j^0$, we also indicate the number $j$ of massless modes and $i$ of tachyons.}\label{tab:eta+IIB2}
  \end{center}
\end{table}

Let us start our comments by mentioning that all de Sitter solutions are perturbatively unstable, i.e.~tachyonic, in agreement with Conjecture 2 of \cite{Andriot:2019wrs}. It is even true for the special solution $s_{55}^+ 18$: this solution was the only known perturbatively stable dS solution without a tachyon in the fields $(\rho,\tau,\sigma_I)$ and we have now proven that it is actually unstable.\footnote{As argued in \cite{Andriot:2021rdy}, this uncommon stability could be related to the fact that the 6d group manifold is in that case non-compact.} This stability does not survive the inclusion of the other scalar fields here. It becomes even strongly unstable, going from $\eta_V \approx 3.8$ to $\eta_V \approx -17.6$. All other solutions are also very unstable with $\eta_V < -1$, most of them however with $|\eta_V| \sim {\cal O}(1)$. This situation is in agreement with the refined de Sitter conjecture of \cite{Garg:2018reu,Ooguri:2018wrx}. Such values of $\eta_V$ are in particular observed for solutions $s_{55}^+ 19, 24, 25$, and $m_{46}^+ 5$. Those four solutions all had $|\eta_V|< 1$ with the restricted set of fields, giving hope that dedicated searches as in \cite{Andriot:2021rdy} could provide viable solutions for a slow-roll cosmological scenario; they now all verify $\eta_V \lesssim -1$. Still, we also observe that for most solutions, the value of $\eta_V$ is not drastically modified when considering all the fields as here. There are a few notable exceptions to the latter, the most impressive change being observed for $m_{5577}^+ 1$ going from $\eta_V \approx -4.75$ to $\eta_V \approx -32.7$. We note that this solution is on a non-compact group manifold \cite{Andriot:2022yyj}. The solution $s_{6666}^+ 1$, which also has a very low value, $\eta_V \approx -20.8$, but does not go through a drastic change, is on a compact manifold. To summarize, all de Sitter solutions are now perturbatively unstable with $\eta_V \leq -1$, i.e.~exhibit strong instabilities.\\

Another important aspect of the instabilities are the field directions of the tachyons. As conjectured in \cite{Danielsson:2012et}, the restricted set of fields $(\rho,\tau,\sigma_I)$ always contains one tachyon, a claim verified in all solutions (except $s_{55}^+ 18$ as mentioned above). Note that the tachyonic direction among these few fields varies, as studied in \cite{Andriot:2021rdy}. Interestingly, we observe here that some solutions have more than one tachyon, meaning that a new one appears, due to the new fields considered. Analysing the directions of the mass matrix eigenvectors, we can determine the fields responsible for the tachyons. Most of the time, a first tachyon (if not the only one) is due to the dilaton and diagonal metric components (sometimes it gets an extra contribution from an off-diagonal metric component): this tachyon is interpreted as the one previously seen, predominantly along $(\rho,\tau,\sigma_I)$. If there is a second tachyon, then the first one typically has the most negative mass squared and is easily distinguished from the other one, which gets further contributions along RR and NSNS axions. This becomes striking in solutions $m_{5577}^+ 9,10$, which have one quasi-massless tachyon predominantly along $C_4$, the other tachyon being the previously known one.

Exceptions to the above general situation go as follows. First, $s_{55}^+20$, $s_{6666}^+ 1$ and $m_{46}^+ 10$ have only one tachyon but with mixed contributions (in particular from axions); at the same time, we note that their value of $\eta_V$ becomes quite low. Turning to the case of two tachyons, the distinction between the two becomes less clear for solutions $s_{55}^+ 18, 19, 22-27$: there the two $m^2 <0$ have similar values, and important contributions of diagonal metric components and axions can be found in both tachyonic directions. This different behaviour of the spectrum may not come as a surprise, since these solutions were all found looking precisely for very specific tachyons \cite{Andriot:2021rdy}. The same phenomenon ($m^2$ values become close, relevant diagonal metric, dilaton and axions contributions in both tachyons) occurs for solutions $m_{46}^+ 3,5, 7-9$ and $m_{55}^+ 1,3$. Finally we note that $s_{6666}^+ 2$ and $m_{5577}^+ 1,2$ exhibit the same mixed contributions in the tachyons, with however two fairly separate negative $m^2$.

It would be interesting to relate these differences in the spectrum to specific features of the solutions. Even though this would deserve more study, we note already that almost all solutions which have more than one tachyon are on non-compact manifolds \cite{Andriot:2022yyj}. The only exceptions are $s_{55}^+ 19, 22-27$ which were however found with non-generic, dedicated searches. This observation should be related to the discussion of massless scalar fields, that we now turn to.\\

Last but not least, we observe the appearance of massless modes; none had been observed before within $(\rho,\tau,\sigma_I)$. Massless modes have been discussed already in Section \ref{sec:flatdir} on flat directions. As indicated in Table \ref{tab:flatdirections}, solution classes $s_{55}$ and $m_{55}$ admit (the only component of) $C_6$ as a flat direction. We recover it here as systematically contributing to the massless modes’ eigenvectors. The other contributions to these eigenvectors in type IIB classes are (some of) the $C_4$ components. For the solutions $s_{55}^+$, there are always 4 massless modes: one is due to $C_6$ and the others to combinations of the 4 $C_4$ components. We see a priori no reason to choose some component of $C_4$ and not the others. Nevertheless, in $s_{55}^+ 12-17, 22-28$, only $C_{4 \ 1356}\,, C_{4 \ 1456}\,, C_{4 \ 2456}$ appear in massless modes, while $C_{4 \ 2356}$ appears (sometimes) in tachyons. Similarly, in $s_{55}^+ 10$, $C_{4 \ 1456}\,, C_{4 \ 2356}\,, C_{4 \ 2456}$ appear in massless modes and $C_{4 \ 1356}$ in tachyons. This is a surprising asymmetry. More generally, we suspect that 3 combinations of $C_4$ components are (non-generic) flat directions of $s_{55}^+$ as discussed in Section \ref{sec:flatdir}, maybe because of $F_5=0$ in all solutions considered. Another surprising observation is the presence of only 3 massless modes in $m_{55}^+$ solutions, while the same fields are present to start with as in $s_{55}^+$. Again, those are due to $C_6$ and combinations of the 4 $C_4$ components. The reduction from 4 to 3 massless modes going from $s_{55}^+$ to $m_{55}^+$ could be due to a different (non-closed) $F_1$-flux in the latter, because of the presence of $D_7$-branes. Finally, in $m_{5577}$, $C_6$ is projected out and only 2 components of $C_4$ remain. The massless mode observed there in some solutions is a combination of these 2 $C_4$ components, but it is also worth noting that some solutions $m_{5577}^+$ do not have any massless mode. It would be interesting to understand why, and more generally, determine combinations of $C_4$ components being flat directions in (subclasses of) type IIB.

Turning to type IIA, we observe the ``T-dual'' behaviour, as already discussed for the number of fields in each class around Table \ref{tab:numberfields}: $s_{66}^+$ and $m_{46}^+$ have 4 massless modes as $s_{55}^+$, while $s_{6666}^+$ solutions have 1 or no massless mode, as $m_{5577}^+$. The class $s_{66}$ allows for 1 $C_5$ component which is a flat direction; the other massless modes are due here to a combination of 4 $C_3$ components (6 are allowed). In $m_{46}$, $C_5$ has 3 components, all flat directions. The remaining massless mode is a combination of 1, 2 or 3 components of $C_3$ among the 4 allowed. Finally in $s_{6666}$, $C_5$ is projected out and $C_3$ has 4 components. The massless mode observed for some solutions is a combination of 2 $C_3$ components. As for type IIB, it would be interesting to prove that combinations of $C_3$ can in type IIA subclasses be flat directions.\\

After this stability analysis, one may wonder whether there is a de Sitter solution which is more promising, phenomenologically, than others. If we stick to basic requirements of compactness of the 6d manifold and the absence of massless mode, we note a surprising (and disappointing) correlation. All solutions of $m_{5577}^+$ with one massless mode are precisely those on a compact manifold, while those without massless mode are on non-compact ones. Similarly, the 4 $m_{55}^+$ solutions with only three massless modes (instead of the four of $s_{55}^+$) are on non-compact manifolds. Regarding type IIA, $s_{6666}^+ 3,4$ which have one massless mode are as well on a compact manifold, while $s_{6666}^+ 2$ which has none is on a non-compact one. The only exception would be $s_{6666}^+ 1$, having no massless mode on a compact manifold. This seems to come however at the cost of a very strong instability: $\eta_V \approx - 20.8$. Finding phenomenologically appealing de Sitter solutions would thus require more efforts. Let us nevertheless give a word of caution on these phenomenological interpretations, because of the distinction between low-energy and consistent truncation mentioned in Section \ref{sec:trunc}: for this warning we refer to the discussion at the end of Section \ref{sec:Mink} on Minkowski solutions. Let us also recall that massless axions could phenomenologically be less problematic, and corrections to our perturbative study may make them massive: see the discussion at the end of Section \ref{sec:flatdir}.

\subsection{Minkowski solutions}\label{sec:Mink}

The mass spectrum of each Minkowski solution is computed with {\tt MSSV}.  We report in Table \ref{tab:eta0II} the number of massless modes and tachyons, comparing them to the number of them within the restricted set of fields $(\rho,\tau,\sigma_I)$. The total number of fields in each class can be found in Table \ref{tab:numberfields} and that of generic flat directions in Table \ref{tab:flatdirections}.

\begin{table}[H]
  \begin{center}
    \begin{tabular}{|c||c|c|c|c|c|c|c|c|c|}
    \hline
class  & \multicolumn{1}{c||}{$s_{55}^0$} & \multicolumn{4}{c||}{$s_{555}^0$} & \multicolumn{2}{c||}{$m_{46}^0$} & \multicolumn{2}{c|}{$m_{466}^0$} \\
\hhline{-||-||----||--||--}
solution & \multicolumn{1}{c||}{1} & 1 & 2 & 3 & \multicolumn{1}{c||}{4} & 1 & \multicolumn{1}{c||}{2}  & 1 & 2 \\
    \hhline{=::=::====::==::==}
$m^2 \leq 0$ \cite{Andriot:2022yyj}  & \multicolumn{1}{c||}{$0^-, 1^0$} & $0^-, 1^0$ & $0^-, 1^0$ & $0^-, 1^0$ & \multicolumn{1}{c||}{$0^-, 1^0$} & $0^-, 1^0$ & \multicolumn{1}{c||}{$0^-, 2^0$} & $0^-, 2^0$ & $0^-, 2^0$ \\
    \hhline{=::=::====::==::==}
$m^2 \leq 0$ &\multicolumn{1}{c||}{$0^-, 7^0$}  & {$2^-, 2^0$} & $0^-, 3^0$ & $0^-, 3^0$ & \multicolumn{1}{c||}{$0^-, 3^0$} & $1^-, 6^0$ & \multicolumn{1}{c||}{$1^-, 7^0$} & $0^-, 3^0$ & $0^-, 3^0$ \\
        \hline
  \multicolumn{10}{c}{}\\
    \hhline{-----~~~~~}
class  & \multicolumn{4}{c|}{$m_{466}^0$} & \multicolumn{5}{}{} \\
\hhline{-----~~~~~}
solution & 3 & 4 & 5 & \multicolumn{1}{c|}{6} & \multicolumn{5}{}{} \\
    \hhline{=::====~~~~~}
$m^2 \leq 0$ \cite{Andriot:2022yyj}  & $0^-, 2^0$ & $0^-, 2^0$ & $0^-, 2^0$ & \multicolumn{1}{c|}{$0^-, 2^0$} & \multicolumn{5}{}{} \\
    \hhline{=::====~~~~~}
$m^2 \leq 0$ & $1^-, 3^0$ & $1^-, 3^0$ & $0^-, 3^0$ & \multicolumn{1}{c|}{$0^-, 3^0$} & \multicolumn{5}{}{} \\
    \hhline{-----~~~~~}
    \end{tabular}
     \caption{Spectrum information for each Minkowski solution in type IIA/B. We indicate by $i^-, j^0$ the number $j$ of massless modes and $i$ of tachyons, first for the fields $(\rho,\tau,\sigma_I)$ as obtained in \cite{Andriot:2022yyj}, and then for the complete set of scalar fields considered here.}\label{tab:eta0II}
 \end{center}
\end{table}

In \cite{Andriot:2022yyj} the Massless Minkowski Conjecture was proposed: it postulates the systematic presence of a massless scalar field among $(\rho,\tau,\sigma_I)$. This was verified for all the solutions we consider here. While including new fields, we observe here the appearance of more massless modes. A first explanation for those are the flat directions due to RR axions, indicated in Table \ref{tab:flatdirections}. This interpretation is perfectly verified for $m_{466}^0$ solutions: while they had 2 massless modes among $(\rho,\tau,\sigma_I)$, we observe here a third one, with field directions in the 3 eigenvectors being purely among the diagonal metric, the dilaton, and $C_5$ (the flat direction). Let us recall that massless modes form a degenerate eigenspace, so the exact field directions of each eigenvector is not a relevant information, the eigenvectors are provided in random combinations. In $m_{46}^0$, eigenvectors are along diagonal metric components and the dilaton as in previous massless modes, together with the 3 $C_5$ components (flat directions) and some of $B_2, C_3$ and off-diagonal metric components. In type IIB, the same holds for $s_{55}^0 1$ with contributions to massless modes from diagonal metric components, the dilaton and $C_6$ (the flat direction), as well as $C_2$ and $C_4$. In these last two examples, it would be interesting, as for de Sitter solutions, to understand why extra axions contribute to massless modes. For $s_{555}^0 1$, the interpretation is again clear, with the same previously known contributions and that of $C_6$, the flat direction. Finally, $s_{555}^0 2-4$ provide an interesting novelty: all three massless modes are along $C_6$, the diagonal metric components and the dilaton. This means that there is one additional massless mode along diagonal metric components and the dilaton, not seen previously with $(\rho,\tau,\sigma_I)$: we verify this by noticing that the eigenvectors distinguish $g_{55}$ and $g_{66}$, while those are not distinguished with $(\rho,\tau,\sigma_I)$ and a set of $O_5/D_5$ along directions 56. The presence of this extra massless mode makes sense, given that the T-dual source configuration in $m_{466}^0$ has 2 massless modes among $(\rho,\tau,\sigma_I)$. This extra mode seems to become tachyonic for $s_{555}^0 1$, a discussion we now turn to.

A surprise in the spectrum of Minkowski solutions are indeed tachyonic directions. In addition to diagonal metric and dilaton contributions, the eigenvectors are along off-diagonal metric and $C_2$ components in $s_{555}^0 1$, and mostly off-diagonal metric components in $m_{46}^0 1,2$ and $m_{466}^0 3,4$. We do not have a clear understanding of their appearance. Let us note however that $s_{555}^0 1$ and $m_{46}^0 1$ have non-compact 6d manifolds, and the compactness has not been established for $m_{46}^0 2$ and $m_{466}^0 3$ \cite{Andriot:2022yyj}; this may allow to discard these solutions. The solution $m_{466}^0 4$ seems however to be on a compact group manifold; maybe the detailed lattice ensuring this compactness should still be investigated.\\

More generally, we conclude that the Massless Minkowski Conjecture is still verified. However, its strong version \cite{Andriot:2022yyj} is in tension with the present results: indeed, it states the absence of 4d tachyonic directions in Minkowski solutions. Nevertheless, one should be careful with the interpretation of this strong version of the conjecture. The latter might indeed be applied more strictly to low-energy effective theories of quantum gravity, while the present consistent truncation is probably not such an effective theory. If a low-energy truncation would keep less modes, the resulting spectrum would be different. In particular, it could avoid the phenomenon of ``space invaders'', where including a priori more massive modes leads to having smaller (or even tachyonic) masses. If this is avoided, the strong version of the Massless Minkowski Conjecture may still hold.

The strong version also refers to other swampland conjectures, and as such could be more sensitive to the connection to string theory. We then note that the validity of the solution $m_{466}^0 4$ as a classical string background has not been tested; this requirement combined with that of the existence of a lattice ensuring the 6d compactness can be challenging \cite{Andriot:2020vlg}. It could then turn out that the present paper does not provide any counter-example to that conjecture; we hope to come back to these matters in future work.

\subsection{Anti-de Sitter solutions}\label{sec:AdS}

We compute for each anti-de Sitter solution the mass spectrum with {\tt MSSV}. We report in Table \ref{tab:eta-II} on the values of the $\eta_V$ parameter, comparing them to the value obtained with the restricted set of fields $(\rho,\tau,\sigma_I)$. We also give the number of massless modes and that of tachyons, obtained with $(\rho,\tau,\sigma_I)$ and obtained here. The total number of fields in each class can be found in Table \ref{tab:numberfields} and the number of generic flat directions in Table \ref{tab:flatdirections}.

\begin{table}[H]
  \begin{center}
  \noindent\makebox[\textwidth]{
    \begin{tabular}{|c||c|c|c|c|c|c|c|c|c|}
    \hline
class & \multicolumn{4}{c||}{$s_{55}^-$} & \multicolumn{5}{c|}{$m_{46}^-$} \\
\hhline{-||----||-----}
solution & 1 & 2 & 3 & \multicolumn{1}{c||}{4} & 1 & 2 & 3 & 4 & 5 \\
    \hhline{=::====::=====}
$\eta_V$ \cite{Andriot:2022yyj} & 0.77850 & $-4$ & $-3.8495$ & \multicolumn{1}{c||}{$-2.4901$} & 1.2531 & 1.5483 & 1.5537 & 1.3004 & 1.2548 \\
    \hhline{=::====::=====}
$\eta_V$ & 1.1436 & 2.5632 & 2.1117 & \multicolumn{1}{c||}{2.7870} & 1.9213 & 1.9873 & 2.0293 & 1.8554 & 2.1590 \\
    \hhline{=::====::=====}
$m^2$ \cite{Andriot:2022yyj} & $1^{\text{{\tiny BF}}}, 0^0$ & $0^{\text{{\tiny BF}}}, 0^0$ & $0^{\text{{\tiny BF}}}, 0^0$ & \multicolumn{1}{c||}{$0^{\text{{\tiny BF}}}, 0^0$} & $1^{\text{{\tiny BF}}}, 0^0$ & $1^{\text{{\tiny BF}}}, 0^0$ & $1^{\text{{\tiny BF}}}, 0^0$ & $1^{\text{{\tiny BF}}}, 0^0$ & $1^{\text{{\tiny BF}}}, 0^0$ \\
    \hhline{=::====::=====}
$m^2$ & $1^{\text{{\tiny BF}}}, 5^0$ & $1^{\text{{\tiny BF}}}, 6^0$ & $1^{\text{{\tiny BF}}}, 6^0$ & \multicolumn{1}{c||}{$1^{\text{{\tiny BF}}}, 6^0$} & $2^{\text{{\tiny BF}}}, 4^0$ & $1^{\text{{\tiny BF}}}, 4^0$ & $1^{\text{{\tiny BF}}}, 5^0$ & $1^{\text{{\tiny BF}}}, 6^0$ & $1^{\text{{\tiny BF}}}, 6^0$ \\
    \hline
    \end{tabular}
    }
     \caption{Spectrum information for each anti-de Sitter solution in type IIA/B. We first provide the value of $\eta_V$ (note the definition and sign convention in \eqref{etaV}) for the fields $(\rho,\tau,\sigma_I)$ obtained in \cite{Andriot:2022yyj}, then the one obtained here with the complete set of scalar fields of the above dimensional reduction. By $i^{\text{{\tiny BF}}}, j^0$, we also indicate the number $j$ of massless modes and $i$ of tachyons (those with $m^2$ below the BF bound).}\label{tab:eta-II}
  \end{center}
\end{table}

Let us first recall that for anti-de Sitter solutions, a 4d perturbatively stable scalar field has a mass $m$ verifying the Breitenlohner-Freedman (BF) bound
\beq
m^2 > -\frac{9}{4 l^2} \quad \Rightarrow \quad \eta_V < \frac{3}{4} \ ,
\eeq
with the anti-de Sitter radius $l$ (see e.g.~\cite[Sec. 3.4.3]{Andriot:2022yyj}). Most solutions were already found unstable within the fields $(\rho,\tau,\sigma_I)$, as can be seen in Table \ref{tab:eta-II}. For these unstable solutions, the addition of fields only makes $\eta_V$ slightly larger, i.e.~the solution more unstable. Looking at the eigenvectors for their tachyonic mode, we do not easily identify the previous tachyon. Diagonal metric components and dilaton always contribute to it, with however off-diagonal metric sometimes contributing, and in addition $C_2$ for $s_{55}^- 1$ and $B_2$ for $m_{46}^- 2,3$.\footnote{We note also for some of these solutions the appearance of other negative $m^2$, however not tachyonic. Their eigenvectors also get mixed contributions.} One of these solutions, $m_{46}^- 1$, gets in addition a second tachyon. Both tachyons there get also very mixed contributions from the various fields.

Three other solutions, $s_{55}^- 2-4$, were found to be perturbatively stable within the fields $(\rho,\tau,\sigma_I)$. These solutions admitted in addition, among these fields, only positive $m^2$. Here, the addition of the new fields generates one tachyon for each of these solutions (and one additional negative $m^2$ for $s_{55}^- 4$). There again, the eigenvectors get very mixed contributions. We conclude that all anti-de Sitter solutions are here found to be unstable (with tachyons partly along axions). Since masses squared below the BF bound are forbidden in supersymmetric anti-de Sitter solutions, we conclude that all the anti-de Sitter solutions above are non-supersymmetric, confirming this suspicion of \cite{Andriot:2022yyj}. The tachyons we found provide perturbative instabilities in agreement with the swampland conjecture of \cite{Ooguri:2016pdq}.\\

Another phenomenon when adding the new fields is the appearance of massless modes. In type IIB, the generic flat direction is $C_6$. In addition, massless modes are along $B_2, C_2,C_4$ for $s_{55}^- 1$ and $B_2, C_4$ for $s_{55}^- 2-4$. In type IIA, the generic flat directions are the 3 components of $C_5$. In addition, massless modes are along $C_3$ for $m_{46}^- 1,2$, $B_2, C_3$ for $m_{46}^- 3$, $B_2, C_1, C_3$ for $m_{46}^- 4,5$. As for de Sitter and Minkowski solutions, it would be interesting to see whether these axionic massless modes are actually flat directions of solution subclasses.

\section{Summary and outlook}\label{sec:conclusion}

In this paper, we derive a 4d theory for compactifications of 10d type II string theory on 6d group manifolds. In particular, we obtain a scalar potential $V$ and the kinetic terms. Our setting includes NSNS- and RR-fluxes as well as (smeared) $O_p$-planes and $D_p$-branes, and the metric fluxes associated to the group manifold. Once implemented numerically in the code {\tt MSSV}, we use this scalar potential to prove that we actually perform a consistent truncation for maximally symmetric spacetimes. We finally analyse the stability of 10d solutions with 4d maximally symmetric spacetimes.

We first describe in Section \ref{sec:trunc} our truncation of the 10d fields, which consists in keeping the left-invariant scalar fields on the group manifold. As discussed there, those are not guaranteed to be the lightest fields (for generic group manifolds) but they are providing a consistent truncation before orientifolding \cite{Cassani:2009ck}. It is expected that the same still holds after including orientifold planes, as shown in various examples. In Section \ref{sec:construnc}, we prove explicitly that our truncation is consistent for all 21 solution classes of \cite{Andriot:2022way}; let us recall that these compactifications include orientifolds. To reach this result, we compare the 10d equations of motion provided by the code {\tt MSSS} \cite{Andriot:2022way} and the 4d equations given by {\tt MSSV} at extrema. Even though a consistent truncation may differ from a low energy truncation, it is sufficient for our purposes when studying the stability of solutions and finding an instability.

Based on this truncation, we derive in Section \ref{sec:dimred} a corresponding 4d theory starting with 10d type II supergravities with $O_p/D_p$ sources. We get in particular the scalar potentials in equations \eqref{VIIA} and \eqref{VIIB}, and compute the scalar kinetic terms. Those allow to define the mass matrix in equation \eqref{massM}, that will provide us with the 4d mass spectrum. This derivation is automated in the code {\tt MSSV}, a Mathematica \cite{Mathematica} notebook presented in Section \ref{sec:tutorial}. This code further computes the mass spectrum for a given solution and related quantities characterising stability. This code is then used in Section \ref{sec:flatdir} to identify generic flat directions in the various solution classes, that would appear as massless modes in the spectrum.\\\vspace{-3mm}

We finally turn in Section \ref{sec:dS}, \ref{sec:Mink} and \ref{sec:AdS} to the stability analysis of 10d solutions with 4d de Sitter, Minkowski or anti-de Sitter spacetimes, respectively. These solutions were obtained in \cite{Andriot:2020wpp, Andriot:2021rdy, Andriot:2022way}, forming a convenient database. Their stability had only been analysed partially on a restricted set of fields denoted $(\rho, \tau, \sigma_I)$ in \cite{Andriot:2022yyj}, and we comment here on the comparison to the stability properties now observed when including all the left-invariant fields. A first result is that all de Sitter and anti-de Sitter solutions at hand are found unstable, despite several candidates found stable within the restricted set of fields. This is in agreement with various conjectures, as discussed in those sections. We also discuss at length the field directions of the tachyons: it is often possible to identify a previously observed tachyon among the fields $(\rho, \tau, \sigma_I)$, while the new fields, e.g.~RR and NSNS axions, contribute to a different one. But in some instances, the contributions of the various fields are more mixed.

We also comment on the numerous massless modes that we observe: some correspond to the generic flat directions, but further contributions from other axions are often noticed. Those may signal further flat directions due to peculiarities of the solutions considered, e.g.~some vanishing flux, that could be viewed as a solution subclass. For Minkowski solutions more specifically, we verify that the Massless Minkowski Conjecture \cite{Andriot:2022yyj} holds, but we also note that checking its strong version is more subtle, and we discuss it. In particular, requiring the compactness of the 6d manifold together with our supergravity solutions being in a classical string regime may not be achieved; this would remove potential counter-examples to this strong version.

In general, correlations are observed between the number of tachyons, of massless modes, or the value of the $\eta_V$ parameter, and whether or not the 6d manifold is compact. (Non)-compactness is one feature of the solutions considered which could explain the differences observed; it would be interesting to identify others.\\

Several questions as well as opportunities are raised after this work, having now the code {\tt MSSV} available; some ideas were already mentioned in the Introduction. To start with, one may consider verifying that the truncation to left-invariant modes on group manifolds and maximally symmetric spacetimes is a consistent truncation, in full generality. By this we mean allowing for any possible $D_p$ source configuration without orientifold. Indeed, within our ansatz, compactifications with orientifolds have all been classified in \cite{Andriot:2022way}, and we checked those here already. In turn, this restricts to anti-de Sitter solutions, according to Maldacena-Nu\~nez no-go theorem \cite{Maldacena:2000mw}: only those do not require orientifolds. While the same procedure could be followed combining the codes {\tt MSSS} and {\tt MSSV}, and we expect this to work, the challenge could be on the amount of equations and variables to consider, larger in absence of orientifold projection. Finally, it would be interesting to go beyond our ansatz with maximally symmetric spacetimes, and verify the consistency of the truncation with time-dependent scalar fields away from the potential extrema; this could amount to time-dependent compactifications, which go beyond the scope of this work.

We suggested in the Introduction to combine the search for 10d solutions performed with {\tt MSSS} and the stability analysis done with {\tt MSSV}. One could even consider using {\tt MSSV} alone, i.e.~the 4d scalar potential, to find new solutions. Their 10d origin would however require to satisfy further constraints, namely the flux Bianchi identities (or tadpole cancellation conditions) and the Jacobi identities on the geometric fluxes. This may still be a useful approach to find interesting critical points of the 4d potential, using new techniques such as gradient descent algorithms. This could be combined with requirements on stability or tachyonic directions, as e.g.~in \cite{Andriot:2021rdy}.

Last but not least, it would be interesting to rewrite the 4d theory obtained here as a 4d (gauged) supergravity. This has been achieved in a variety of examples (see Section \ref{sec:trunc}). We mentioned there in particular the approach using $\rm{SU}(3)\times \rm{SU}(3)$ structures, allowing to reach a 4d ${\cal N}=2$ supergravity, or ${\cal N}=1$ with an orientifold projection. While this formalism is very appealing, it seems unfortunately not to apply to some of our solution classes, where we reach ${\cal N}=1$ via the specific placement of a $D_p$-brane (and not an $O_p$-plane). It is for instance the case for the 28 de Sitter solutions of $s_{55}^+$, where the source configuration ($O_{5}$ along directions 12, 34, $D_5$ along 56) preserves ${\cal N}=1$ supersymmetry in 4d. Formulating the corresponding 4d theory as an ${\cal N}=1$ supergravity seems then challenging, but we hope to come back to this question in future work.

\section*{Acknowledgements}

We would like to thank D.~Tsimpis for helpful discussions. The work of M.R.~and T.W.~is supported in part by the NSF grant PHY-2013988. M.R.~acknowledges the support of the Dr.~Hyo Sang Lee Graduate Fellowship from the College of Arts and Sciences at Lehigh University.

\bibliographystyle{JHEP}

\providecommand{\href}[2]{#2}\begingroup\raggedright\endgroup

\end{document}